\documentclass{article}
\usepackage{amssymb,amsmath,latexsym}
\usepackage{graphicx}

\begin{document}

\title{ {\large Can Specific THz Fields Induce Collective Base-Flipping in DNA?}\\
{A Stochastic Averaging and Resonant Enhancement Investigation  \\ Based on a New Mesoscopic Model }
}
\date{March 15, 2024}
\author{Wang Sang Koon\footnote{Corresponding Author.  koon@caltech.edu. Control and Dynamical Systems,  California Institute of Technology, Pasadena, California, 91125}, 
Houman Owhadi\footnote{owhadi@caltech.edu. Applied and Computational Mathematics and Control and Dynamical Systems, California Institute of Technology, Pasadena, California, 91125}, 
Molei Tao\footnote{mtao@gatech.edu. School of Mathematics, Georgia Institute of Technology, Atlanta, Georgia, 30332}, 
Tomohiro Yanao\footnote{yanao@waseda.jp. Department of Applied Mechanics and Aerospace Engineering, Waseda University, Tokyo, Japan}
}
\maketitle
\begin{abstract}

We study the metastability, internal frequencies, activation mechanism, energy transfer, and the collective base-flipping in a mesoscopic DNA via resonance with  specific electric fields. Our new mesoscopic DNA model takes into account  not only the issues of helicity and the coupling of an electric field with the base dipole moments, but also includes  environmental effects such as fluid viscosity and thermal noise.  And all the parameter values are chosen to best represent the typical values for the opening and closing dynamics of a DNA. Our study shows that while the mesocopic DNA is metastable and robust to environmental effects, it is vulnerable  to certain frequencies that could be targeted by  specific THz fields for triggering its collective base-flipping dynamics and causing large amplitude separation of base pairs. 
Based on applying Freidlin-Wentzell method of stochastic averaging and the newly developed theory of resonant enhancement to our mesoscopic DNA model, our semi-analytic  estimates show that the required  fields should be THz fields with frequencies around 0.28 THz and with  amplitudes in the order of 450 kV/cm. These estimates compare well with the experimental data of Titova et al.,  which have demonstrated that they could affect the function of DNA in human skin tissues by THz pulses with frequencies around 0.5 THz and  with a peak electric field at 220 kV/cm.  Moreover, our estimates also conform to a number of other experimental results which appeared in the last couple years.  

{\bf 
Our study shows that while the mesocopic DNA is metastable and robust to environmental effects, it is vulnerable  to certain frequencies that could be targeted by  specific THz fields for triggering its collective base-flipping dynamics and causing large amplitude separation of base pairs. 
Though more work may still be needed to take into consideration the issue of modeling the pulses,  we believe that, even as it stands right now, our new mesoscopic DNA model and semi-analytical estimates have suggested resonance as a possible mechanism  for the experimental results mentioned above, and hopefully will inspire new experimental designs that may settle this question.  Furthermore, our semi-analytical methods may be useful in studying the  metastable rate and its resonant enhancement  for  chemical as well as other bio-molecular systems.}

\end{abstract}



\section{Introduction}

We study the metastability, internal frequencies, activation mechanism, energy transfer, and the collective base-flipping in a mesoscopic DNA via resonance with specific electric fields. Our study shows that while the mesocopic DNA is metastable and robust to environmental effects, it is vulnerable  to certain frequencies that could be targeted by  specific THz fields for triggering its collective base-flipping dynamics and causing large amplitude separation of base pairs.  

THz radiation (0.1 THz to 10 THz) lies between the infrared and microwave regions and is  non-ionizing. And the mechanisms by which it interacts with cells and tissues are fundamentally different from those involved in interactions of living matter with high-energy ionizing radiation that cause damage by directly breaking covalent bonds in DNA and other biomolecules \cite{TiAy13c}.
 
Since THz field is a frontier area for research in physics, chemistry,  biology, material science and medicine, their impact on DNA dynamics is an extremely important question.  In their study of DNA breathing dynamics in the presence of a THz field -- based on the PBD model \cite{DaPe93} with a drag term and a harmonic driving force, 
Alexandrov et al. \cite{AlGe10} argued that a specific THz radiation exposure may significantly affect the natural dynamics of DNA, and thereby influence intricate molecular  processes involved in gene expression and DNA replication. Lots of controversy follow:  some considered their paper excellent while others saw it as "physically unrealistic" \cite{Sw11, MaMi22}. 

Titova el al. \cite{TiAy13a, TiAy13b, TiAy13c} brought the interest to another level.  They demonstrated in vitro, and in vivo that intense (220 kV/cm), picosecond, THz pulses with 1 KHz repetition rate can elicit cellular molecular changes in exposed skin cells and tissues in absence of thermal effects.  They observed that exposure to intense THz pulses (i) leads to a significant induction of H2AX phosphorylation which indicates DNA damage, and at the same time an increase in several proteins which suggest that DNA damage repair mechanisms are quickly activated, (ii) causes favorable changes in the expression of multiple genes implicated in inflammatory skin diseases and  skin cancers which suggests potential therapeutic applications of intense THz pulses.  Moreover, they pointed out that (iii) the low average power of THz pulse sources in their experiment results in biologically insignificant temperature increases of only fractions of a degree at most, (iv) but the high energy density per pulse produces peak powers and corresponding electric fields that can be extremely high and it is likely that these high peak electric fields are responsible for the observed THz-pulse-driven cellular effects \footnote{See Section  \ref{Titova} for more details on the experiments of Titova et al. and the discussion of the issue of  waves vs pulses. }. 

In their papers, they also mentioned the work of Alexandrov et al. in a favorable light and hoped to understand better the mechanisms, and the specific requirements for the intensity and frequency of the radiation.

\subsection{A New DNA Model with Environmental Effects and Radiation}

Building on our earlier work on DNA \cite{KoOw13}, the insights from this controversy,  and especially the non-thermal experimental results of Titova et al., we develop a  new mesoscopic DNA model in Section \ref{Model} of this paper that takes into account not only the issues of helicity and the coupling of an  electric field with the base dipole moments, but also includes the environmental effects such as  fluid viscosity and thermal noise \cite{EnCa80, Yo83, Ya04, Zh89, Me06}.  Moreover, all the parameter values are chosen to best represent the typical values for the opening and closing dynamics of a DNA  \cite{CaDe11, DeDe08}.

We start with a mesocopic DNA in the fluid with thermal noise,  but disregard the additional thermal effects by ignoring radiation-fluid interactions (i.e., in the non-thermal regime), in order to  focus our attention mainly on the effects of radiation on the DNA via coupling the electric field  with the DNA bases.  The new mesoscopic DNA model allows us to employ Freidlin-Wentzell theory of stochastic averaging \cite{FrWen12, FrWeb98, FrWeb01, Fr98, StVa97}  and the newly developed theory of resonant  enhancement \cite{ChTa22, DyRa97, SmDy97, DyGo01}. Our study in Section \ref{FW} and Section \ref{CT} shows that while the mesocopic DNA is metastable and robust to environmental effects, it is vulnerable  to certain frequencies that could be targetted by  specific THz fields for triggering its collective base-flipping dynamics and causing large amplitude separation of base pairs.

\subsection{The Mesocopic DNA is Robust Against  Noise}

To determine whether the mesocopic DNA is robust to noise and will survive in a normal environment without radiation, we  estimate its mean life span, $T_{DNA}$, where the base pairs  have not gone through a collective base-flipping and  have not had a large amplitude separation, due to environmental random perturbations. 

For our study, $T_{DNA}$ is nothing but the mean first passage time (MFPT) for the 2N-dimensional diffusion process defined by the kinetic Langevin system  of our model, namely, Eq.(\ref{EOM_PR_dimensionless})  with ${\mathcal E}=0$, i.e., without radiation. In order to estimate this MFPT$_{\rm 2N}$, we proceed as follows in Section \ref{FW} : (1) we set up the MFPT equation for this $2N$-dimensional Hamiltonian diffusion process with weak noise; (2) we use Freidlin-Wentzell method of Hamiltonian stochastic averaging to reduce the above equation to  a MFPT equation for an 1D energy diffusion process defined on a graph;  (3) we simplify and use Monte Carlo numerical methods to obtain the drift and diffusion coefficients for this stochastic averaged  equation; and (4) we solve this stochastic averaged equation and obtain an approximation of the MFPT$_{\rm 2N}$ for the full 2N-dimensional diffusion process. As it turns out, the mean first passage time of our mesoscopic DNA, MFPT$_{\rm 2N}$, is equal to $6.1487\times 10^{18}\,\, {\rm picoseconds}$ from our semi-analytical computation.

Hence, $T_{DNA}=6.1487\times 10^{18}\,\, {\rm ps}$ which is about 10 weeks.  Since the mean life time of a skin cell is about 2 to 4 weeks, and $T_{DNA}>T_{cell}$,  we can reasonably conclude that the mesocopic DNA  is robust against noise and will survive in a normal environment without radiation. 

\subsection{But It is Vulnerable to  Specific THz Fields }

After showing that the mesoscopic DNA is robust against noise, we try to quantify the requirements for specific electric fields that can trigger its collective base-flipping dynamics and cause large amplitude separation of base pairs. We use the newly developed theory of resonant enhancement of the rate of metastable transition \cite{ChTa22}, which has been built on the work of large deviation theory \cite{FrWen12} and the results of  researches in physics \cite{DyRa97, SmDy97, DyGo01}.

In Section \ref{CT}, we first briefly summarize the main results of this new theory of resonant enhancement and then apply it to our mesoscopic DNA model.  And we find the requirements for the specific electric fields in the following steps: (1) we obtain numerically the most probable path (MPP) which is essentially the heteroclinic connection between the metastable equilibrium point at the bottom of the left potential well and  the rank-one saddle;  (2) we derive analytically the formula for the work done by the electric field along the MPP that winds from the bottom to the top of the potential well;  (3) we use this formula to find, via numerical methods, the requirements for the specific electric  fields  and conclude that the mesoscopic DNA is vulnerable to  THz fields with frequencies around 0.28 THz and with amplitudes in the order of 450 kV/cm. 

These estimates compare well with the experimental data of Titova et al.  \cite{TiAy13a, TiAy13b, TiAy13c}, which have demonstrated that they could affect the function of DNA in human skin tissues by THz pulses with frequencies around 0.5 THz and  with a peak electric field at 220 kV/cm.  Moreover, our estimates also conform to a number of other experimental works \cite{SiIl21, HoPu21, AbEr21} which appeared in the last couple years.  


\subsection{The Need to Model THz Pulses}

But more study is needed to take into consideration the issue of modeling the pulses. 
In the experiments of Titova et al., they skillfully used intense THz pulses with low repetition rate to obtain  a sufficiently high peak electric field and to affect the DNA function in a non-thermal setting. In our present study, we try to theoretically investigate  the mechanisms and requirements for such radiation using the existing tools in the field of control of stochastic mechanical systems. By ignoring the additional radiation-fluid interaction, we get ourselves into the non-thermal setting, minimize the impacts of the issue of waves vs pulses, and enable us to obtain some interesting and important results.   And we wish that future research will extend the present theory of resonant enhancement  to the case where the radiation are pulses.  However, even as it stands right now, our new DNA model and semi-analytical estimates may have suggested resonance as a possible mechanism for these experimental results and hopefully will inspire new experimental designs that may settle this issue.  

Moreover, we believe that our semi-analytical methods may be useful in studying the metastable rate and its enhancement  for  chemical as well as other bio-molecular systems.


\section{A New DNA Model with Environmental Effects and Radiation}\label{Model}

\begin{figure}[ht]
\begin{center}
\includegraphics[width=4.7in]{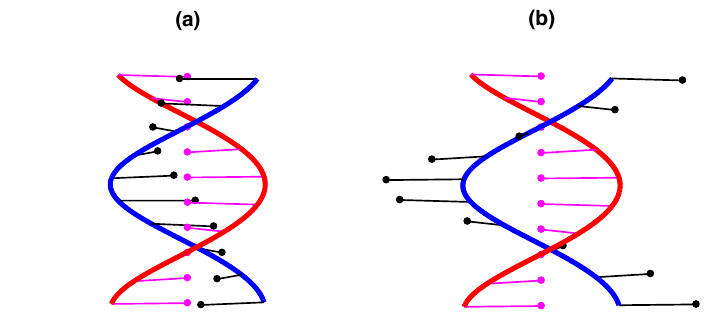}  
\end{center}
\caption{\label{DNAmodel} {\footnotesize
A model of DNA opening dynamics  with 10 pendula (for illustration only, but $N=30$ throughout  this paper).  The work of Yomosa  has shown us how to generalize the model of Englander et. al.  to a helical model simply by twisting both strands forward by 36$^{\tiny o}$.
The system has two  global stable equilibria, achieved when all the pendula (in black color) have the same angular displacements $\theta _n=\theta _e$ and each is positioned at the equilibria of a single pendulum.  Figure (a) shows the case where $\theta_e=-2.6906$ (left equilibrium point in Figure (\ref{DNA_PS})).  
(b) The system also has a rank one saddle at $\theta _n=0$  (and $p_{\theta_n}$=0) where all pendula (in black color) are at the outward position. 
For a small energy, the pendula are librating near one of the global stable equilibria  where all angles $\theta _n$ are the same and equal to $\theta _e$.  For a large enough energy,  the pendula will  move collectively from one energy basin to the other and  flip across the rank one saddle.}}
\end{figure}

\subsection{The Torsional Model of DNA Opening Dynamics} 

\begin{figure}[ht]
\begin{center}
\includegraphics[width=3in]{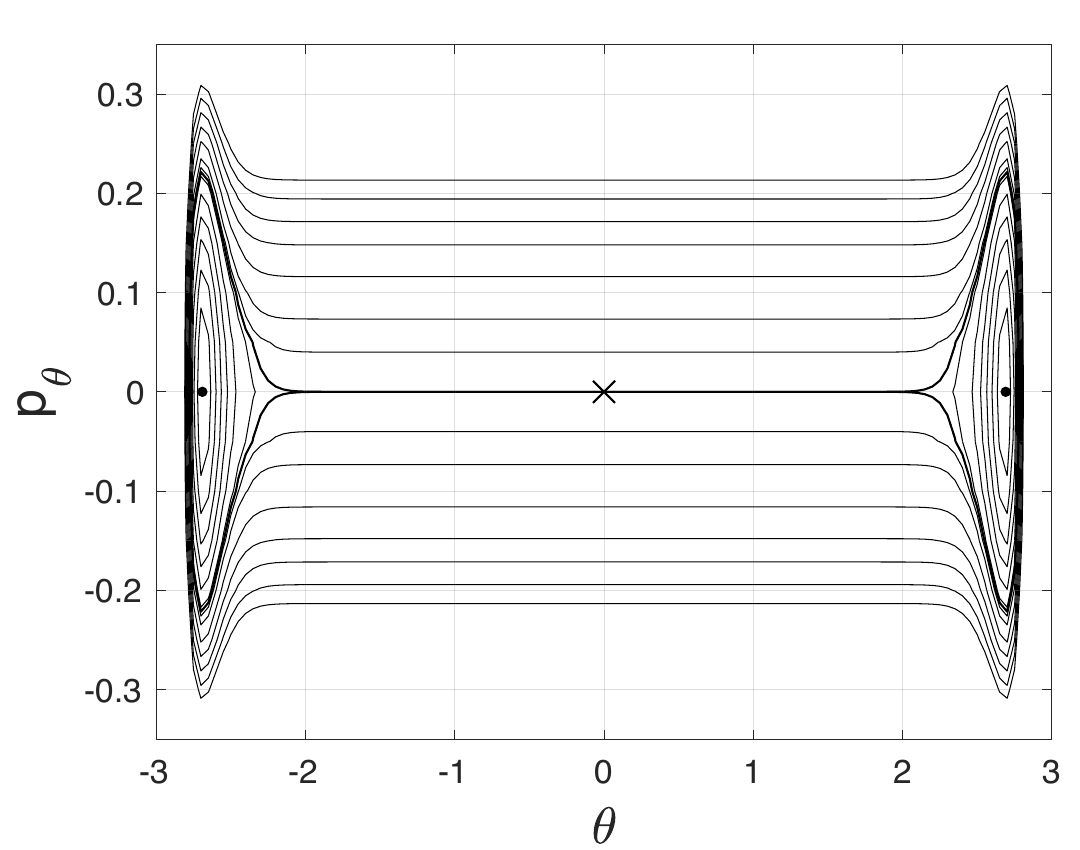} 
\end{center}
\caption{\label{DNA_PS} {\footnotesize
The phase space $(\theta ,p_\theta )$ of a single pendulum in a Morse potential and without torsional coupling with its nearest neighbors. It has two stable equilibria  and a saddle. The curve containing the saddle at $(0, 0)$ is a homoclinic orbit  that separates  two types of motion, namely, the libration near the equilibria $(\theta _e,0)=(\pm 2.6906,0)$, and the flipping across the saddle $(0,0)$.}}
\end{figure}

The torsional  model \cite{EnCa80, Yo83}  is a chain of equivalent pendula (in black color) that rotate about the axis of a fixed backbone with an angle $\theta _n$ measured from the outward position. The pendula interact with nearest neighbors along the backbone through harmonic torsional coupling, and with pendula (in pink color) on the opposing immobilized strand through a Morse potential that has two stable equilibria and a saddle. See Figure (\ref{DNAmodel}) and Figure (\ref{DNA_PS}).

\subsubsection{Brief Remarks on the History of the Torsional Model} 

In a pioneering paper  \cite{EnCa80}, biophysicists Englander et al. first modeled the DNA as a double chain of coupled pendula, and employed  it to investigate the base pair opening and  motion of transcription bubble. Many researchers  built on this torsional model.  Yomosa \cite{Yo83} generalized it to a helical model by twisting both strands forward by 36$^{\tiny o}.$ Russian biophysicist Yakushevich extended it  in a series of  papers, making it an excellent first model in exploring the DNA opening dynamics and transcription \cite{Ya04}.  More recently, Mezic, as well as others, used it in studying the structural activation of large-amplitude collective motions of the bases \cite{Me06, DuMe09, EiMe10, KoOw13}.

In a related but different direction, other researchers  studied the  models for the radial stretching motions of DNA molecule, which are related in particular to DNA thermal denaturation. The main line of research in this direction followed the formulation of Peyrard-Bishop-Dauxois model  \cite{PeBi89, DaPe93}, which Alexandrov et al. used in exploring the effects of a THz field on the DNA breathing dynamics \cite{AlGe10} . 

\subsubsection{The Hamiltonian for the DNA Opening Dynamics}

The Hamiltonian that describes the motion of these $N$ coupled pendula is given by
\begin{equation}\label{DNAmodeleq0}
H_0(\theta ,P_ \theta )=\sum _{n=1}^N
\left[\frac{P_{\theta n}^2}{2 I}+
\frac{1}{2}S(\theta _n-\theta _{n-1})^2+D
\left( e^{-a_d \{ l(1+\cos \theta _n)-x_0 \} }-1\right)^2\right]
\end{equation}
with periodic boundary condition, $\theta _{0}=\theta _N$. $N=30$ for this study. The first term is the kinetic energy terms of the $N$-pendula. The second term is the torsional coupling terms. The third term is the Morse potential terms, which model the hydrogen bonds of the respective DNA base pairs. In Eq.\,(\ref{DNAmodeleq0}), $P_{\theta n} =I (d\theta_n/ dt') $ is the generalized momentum   conjugate to $\theta_n$ where $t'$ is the natural time of the model.  All the parameter values are chosen to best represent the typical values for the opening and closing dynamics of DNA  \cite{CaDe11, DeDe08} and are given as follows:
\begin{enumerate}
\item
The parameter $I$ represents the moment of inertia of each pendulum  and is given by
$I=18600\times 1.67\times 10^{-47} \, \,  {\rm kg\cdot m^2}=310.62 \,\,  {\rm pN\cdot nm\cdot ps^2}$.
\item
The parameter $S$ determines the strength of the nearest neighbor coupling and is given by 
$S=2 \,\, {\rm eV }=320.4 \,\, {\rm pN\cdot nm}$.
\item 
The parameter $D$ determines the strength of the Morse potential and is given by
$D=0.05 \, \, eV=8.01\,\, {\rm pN\cdot nm}$.
\item
$x_0, a_d$, and $h$ represent the equilibrium distance, the width of the Morse potential, and the length of each pendulum, respectively.  Their values are given by $x_0 = 0.1$ nm,  $a_d = 19$ nm$^{-1}$, and $l=1$ nm.
\end{enumerate}

After dividing both sides of  Eq.\,(\ref{DNAmodeleq0}) by $S$, the Hamiltonian can be non-dimensionalized as
\begin{equation}\label{DNAmodeleq}
H(\theta ,p_ \theta )=\sum _{n=1}^N \left[\frac{1}{2}p_{\theta n}^2+ \frac{1}{2}(\theta _n-\theta _{n-1})^2+\kappa\left( e^{-a(1+\cos \theta _n-d_0 ) }-1\right)^2\right]
\end{equation}
where $p _{\theta n} \equiv d\theta _n/ dt$ is the dimensionless momentum defined with respect to the dimensionless time $t= \sqrt{S/I }\,\, t'$. Thus, in this study, one unit time ($t=1$) corresponds to $t' = \sqrt{I/S} =$0.9846 ps.  In Eq.(\ref{DNAmodeleq}), the dimensionless amplitude of the Morse potential $\kappa $ is a small parameter and is given by $\kappa = D/S= 0.025$. We have also introduced the dimensionless decaying coefficient of Morse potential $a \equiv a_d l = 19$, and the dimensionless equilibrium distance $d_0 \equiv a_d x_0/a = 0.1$.

\subsubsection{Libration and Collective Base-Flipping} 

Before studying the dynamics of this N-pendula torsional model, it is instructive to look at  a single pendulum in a Morse potential without coupling.  Figure (\ref{DNA_PS})  shows the phase space of such single pendulum. It has two stable equilibria  and a saddle.  The curve containing the saddle $(0,0)$ is a homoclinic orbit  that separates  two types of motion, namely, the oscillation near the equilibria $(\theta _e,0)=(\pm 2.6904,0)$, and the flipping across the saddle $(0,0)$.  The N-coupled pendula have similar but much more complicated behaviors.  First, the system has two  global stable equilibria, achieved when all the pendula have the same angular displacements $\theta _n=\theta _e$ (thus nullifying the nearest neighbors coupling) and each is positioned at the equilibria of a single pendulum.  It also has a rank one saddle at $\theta _n=0$  (and $p_{\theta_n}$=0) where all pendula are at the outward position.  For a small energy, the pendula are librating near one of the global stable equilibria  where all angles $\theta _n$ are the same and equal to $\theta _e$.  For a large enough energy,  the pendula will  move collectively from one energy basin to the other and to flip across the rank one saddle. Numerical simulations show that  the minimum activation energy for flipping depends on the way how the energy is injected  \cite{DuMe09, EiMe10, KoOw13}.

\subsection{Equations of Motion for the New DNA Model } 

\begin{figure}[htb]
\begin{center}
\includegraphics[width=4.8in]{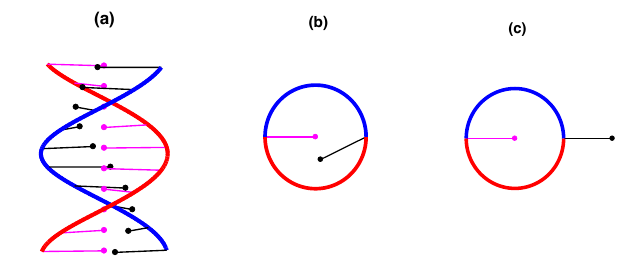}
\end{center}
\caption{\label{EMF} {\footnotesize
Figure (a) shows the DNA model at its left global equilibrium point where $\theta_n=\theta_e=-2.6906$. 
Figure (b) shows the positions of the first base pair at the global left equilibrium point, with pink one immobilized and black one at $\theta _e=-2.6906$. 
The work of Zhang provides us with the idea of coupling an  electric field with the dipole moments of DNA bases.  If we assume that the electric field is perpendicular to the DNA central axis and is directed to the inverted direction of the first base shown in Figure (c), then the moment of force exerted on the $n$-th base can be written as $T_n=-P_n\sin (\theta _n+(n-1)\pi/5)\,\,E\cos \Omega_0 t' $.
}}
\end{figure}

The new DNA model, as anticipated  in our previous paper  \cite{KoOw13}, will take into account the issues of  (i) helicity, (ii) coupling of an electric field with the base dipole moments, and (iii) environmental effects of fluid viscosity and thermal noise.   It will allow us to explore the possibility, the requirement, and the mechanism for an electric field to trigger its collective base-flipping dynamics,
cause large amplitude separation of base pairs, and disrupt the function of DNA \footnote{We will consider the coupling of an electric field with the base dipole moments, but will ignore the additional temperature effect induced by the radiation-fluid interaction.
See Section \ref{Titova} for more details.  }.

\subsubsection{The Coupling of an Electric Field with the Bases of a DNA} 

As mentioned earlier, the work of Yomosa \cite{Yo83} has showed us how to generalize the model of Englander et al. to a helical model simply by twisting both strands forward by 36$^{\tiny o}.$  And the work of Zhang \cite{Zh89} provides us with the idea of coupling an  electric field with the dipole moments of DNA bases.  If we assume that the electric field is perpendicular to the DNA central axis and is directed to the inverted direction of the first base, then the moment of force exerted on the  $n$-th base can be written as 
\begin{equation}\label{Tn}
T_n=-P_n\sin (\theta _n+(n-1)\pi/5)\,\,
E\cos \Omega_0 t'
\end{equation}
where $E,\Omega_0$ are the amplitude and the angular  frequency of the electric field, and  $P_n$ is the dipole moment of the $n$-th base.  The helical geometry requires the $\pi/5 $ twisting.  See Figure (\ref{EMF}). Below, for convenience, we will ignore the negative sign  in Eq. (\ref{Tn}) simply by modifying the electric field by half of a cycle.  

Moreover, since the model is homogeneous in this paper, we will assume that $P_n=P$ from now on where $P$ is the average dipole moment taken from the work of Zhang. As for the issue of inhomogeneity, our initial computations have convinced us  that the frequencies and amplitudes of the required fields for triggering the collective base-flipping  are definitely sequence dependent (A,T,G,C), but we will leave them to our next paper.  

\subsubsection{The Inclusion of Environmental Effects }\label{Formula} 

After including environmental effects \cite{BoOw10, LeMa15, Ot96},  the new model can be seen as a  periodically-driven kinetic Langevin system for the torsional model:
\begin{eqnarray}\label{EOM_PR}
d\theta _n &=& I^{-1}P_{\theta n} \, dt' \nonumber \\
dP_{\theta n}&=& S(\theta _{n+1} -2\theta _n+\theta _{n-1}) \, dt' 
-DU_0'(\theta _n) \, dt' 
 -\gamma I^{-1}P_{\theta n} \, dt'  \nonumber \\
&& + F\sin(\theta _n+(n-1)\pi/5) (\cos\Omega _0\,t') \, dt'
+ \sqrt{2\beta_0 ^{-1} \gamma } \, dW
\end{eqnarray}
where $n=1,...,N$, $\theta _{0}=\theta _N$, $U_0' $ is the derivative of  the Morse potential $U_0 $.  It can be non-dimensionalized as 
\begin{eqnarray}\label{EOM_PR_dimensionless}
d\theta _n &=& p_{\theta n} \, dt \nonumber \\
dp_{\theta n}&=& (\theta _{n+1} -2\theta _n+\theta _{n-1}) \, dt 
-\kappa U'(\theta _n) \, dt 
-\nu \kappa p_{\theta n} \, dt \nonumber \\
&&
+ {\mathcal E}\kappa \sin(\theta _n+(n-1)\pi/5) (\cos\Omega t) \, dt
+ \sqrt{2\beta  ^{-1} ( \nu \kappa) } \, dW
\end{eqnarray}
where $U'$ is the derivative of the non-dimensional Morse potential $U$.  As for the parameters of these two systems, they are related as follows:
\begin{enumerate}
\item
$\gamma I^{-1}$ is the frictional coefficient which models the damping  of the collective base rotation inside the DNA by the surrounding fluid molecules: $\nu \kappa =\gamma I^{-1}\sqrt{I/S}$.
Usually, it is defined via the time of relaxation and is in the order of $100$ ps \cite{SiRa13, Ya04}. In this work, we set $t'_r=150 \, {\rm ps}$. Hence
\begin{equation}\label{under-damped}
\nu\kappa =\frac{1}{150\, ({\rm ps})}(\sqrt{I/S})=0.006564
\end{equation}
and the system is underdamped.
\item
$\beta _0 ^{-1}$ is the Boltzmann temperature which models the thermal energy gain: $k_B T= \beta _0^{-1}=S\beta  ^{-1}$ where $k_B$ is the Boltzmann constant and $T$ is the absolute temperature.
For body temperature, T= 273 + 37 = 310 K, and $\beta^{-1}$ is given by 
\[
\beta  ^{-1}=k_BT/S=4.2811\times 10^{-21}{\rm J}\times 6.242\times 10^{18} {\rm (eV)/J/(2 eV)}
=0.01336
\]
\item
$F=P\cdot E$ is the  amplitude  of the excitation which models  the moment of force exerted on DNA base by the electric field: $PE={\mathcal E} \kappa S$.  For average dipole moment,  we set $P=5.3$ db \cite{Zh89}.  And the electric field amplitude is given by
\begin{equation}\label{E}
E=\frac{8.01\times {\mathcal E}}{0.3356\times 10^{-8}\times P}\,\,({\rm V/m})
=4503.3\times {\mathcal E}\,\,({\rm kV/cm}) 
\end{equation}
Here, It is interesting to note that if ${\mathcal E}$ is in the order of $0.1$, then $E$ will be in the order of $400$ kV/cm as in the experiments of Titova et al. \cite{TiAy13a, TiAy13b, TiAy13c}. See {Section \ref{Titova}}.
\item
$\Omega_0$ and $\Omega $ are the angular frequencies of excitation in the respective system: $\Omega _0=\Omega \sqrt{S/I}$ (radian/ps). Hence, the resonant frequency 
\begin{equation}\label{Omega}
\Omega _0/2\pi=1.0156\times \Omega/2\pi \,\, {\rm(rev/ps=THz)}.
\end{equation}
If the non-dimensional $\Omega$ is in the order of $1$, then the resonant frequency will be in the order of $0.1$ THz and can be considered as a THz field. 
\end{enumerate}

\subsection{Robust Against Noise But Vulnerable to Specific THz Fields? }

After developing a new helical DNA model that takes into consideration all the relevant issues such as fluid and radiation,  and selecting the parameter values that  best represent the typical data for the opening and closing dynamics of a DNA, we are ready to explore the possibility, the requirement, and the mechanism for a electric field to trigger its collective base-flipping dynamics, causing large amplitude separation of base pairs, and disrupting the function of DNA.

But before we start, we need to first check that our mesoscipic DNA is robust against noise when there is no radiation, i.e., when ${\mathcal E}=0$.  Since the forceless system is a kinetic Langevin system with weak noise, we can quantify the metastability of its normal state where the pendula librate near its equilibrium point,  via Freidlin-Wentzell method of  Hamiltonian stochastic averaging.  And we need to determine whether the mean life span of our mesoscopic DNA -- where the base pairs have not gone through a collective base-flipping and have not had a large separation of base pairs due to environmental perturbations, is sufficiently long enough so that we can have confidence  that it  will survive in a normal environment without radiation.  

In Section \ref{FW}, we will answer the question: "is the mesoscopic DNA robust against noise?"  In Section \ref{CT}, we will deal with the issue:  "is it vulnerable to specific THz fields?".



\section{The Mesocopic DNA is Robust Against  Noise}\label{FW}

\begin{figure}[htp]
\begin{center} 
\includegraphics[width=4.8in]{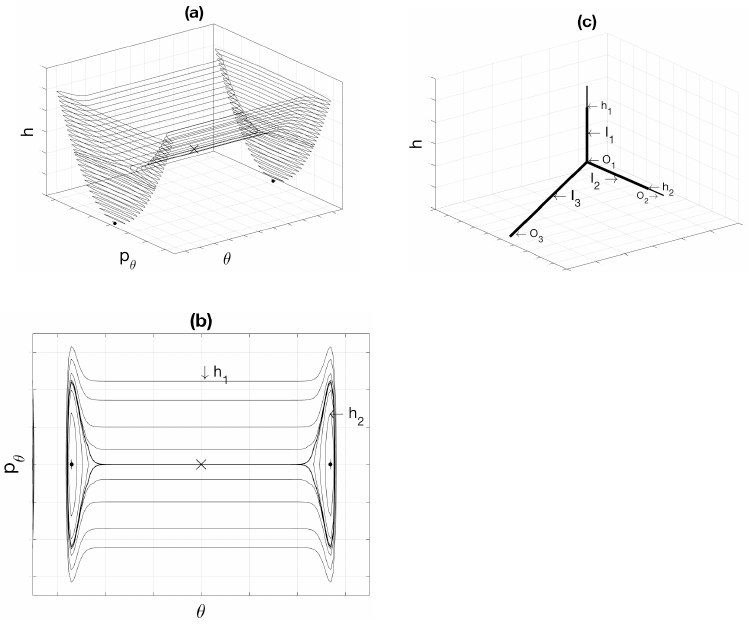} 
\end{center}
\caption{\label{Graph2} {\footnotesize
These figures are  for conceptual illustrations only.
(a) A $(2N+1)$-dimensional Hamiltonian contour plot with Figure (b) as its corresponding $2N$-dimensional version.  See references  \cite{FrWeb01, FrWen12} for the geometry of Hamiltonian systems with many degrees of freedom.
(b) $G$ is a bounded region in the 2N-dimensional space $(\theta, p_\theta)$ with the boundary $\partial G=\partial G_2 \bigcup \partial G_1$,  where $\partial G_2=\{(\theta, p_\theta): H(\theta,p_\theta)=h_2 \,\,\,\, \& \,\,\,\, \theta>0\}$ is a energy level set blocking 
away a small neighborhood of the bottom of the right well.
$\partial G_1=\{(\theta, p_\theta): H(\theta,p_\theta)=h_1\}$ is another appropriately  chosen energy level set with its energy higher than the energy of the rank-one saddle.
(c) Graph $\Gamma$ is homeomorphic to the set of connected components
of level sets of $H(\theta, p_\theta)$.  $\Gamma $ consists 3 edges, $I_1,I_2,I_3$,  with $I_3, I_2$ corresponding to those in the left and right well respectively, and with $I_1$ corresponding to those above the saddle.
$\Gamma $ also has 3 vertices, with $O_2,O_3$ corresponding to two equilibria,  and $O_1$ corresponding to the energy level set containing the rank-one saddle.
}}
\end{figure}

To determine whether the mesocopic DNA is robust to noise and will survive in a normal environment without radiation, we  estimate its mean life span, $T_{DNA}$, due to environmental random perturbations, where the base pairs  have not gone through a collective base-flipping and  have not had a large amplitude separation. For the rest of the paper, we will call it the mean life span of collective base-flipping and use it to mark the life span of our mesoscopic DNA. 
 
For our study, $T_{DNA}$ is nothing but the mean first passage time (MFPT) for the 2N-dimensional diffusion process defined by the kinetic Langevin system of our model, Eq.\,(\ref{EOM_PR_dimensionless}) with ${\mathcal E}=0$, i.e., without radiation.  In order to estimate this MFPT$_{\rm 2N}$, we will proceed as follows: (1) we will set up the MFPT equation for this $2N$-dimensional Hamiltonian diffusion process with weak noise; (2) we will use the Freidlin-Wentzell  method of Hamiltonian stochastic averaging to reduce the above equation to  a MFPT equation for an 1D energy diffusion process defined on a graph;  (3) we will simplify and use Monte Carlo numerical methods to obtain the drift and diffusion coefficients for this stochastic averaged  equation;  and (4) we will solve this stochastic averaged equation and obtain an approximation of the MFPT$_{\rm 2N}$ for the full 2N-dimensional diffusion process.

As it turns out, our semi-analytical estimations in this section will show that the mean life span of collective base-flipping of the mesoscopic DNA, $T_{DNA}$, is equal to $6.1487\times 10^{18}\,\, {\rm ps}$,  which is about 10 weeks long.  Since the life time of a skin cell is between 2 to 4 weeks, and $T_{DNA}>T_{cell}$,  we can reasonably conclude that the mesocopic DNA  is robust against noise and will survive in a normal environment without radiation. Moreover, a closer analysis of these estimates will also allow us to remark that FW theory is essential in studying the MFPT and its corresponding metastable rate.

\subsection{MFPT for a 2N-dim. Hamiltonian Diffusion  with Weak Noise  } 

Let $G$ be a bounded region in $R^{2N}$ with the smooth boundary $\partial G=\partial G_2 \bigcup \partial G_1$,  where $\partial G_2$ is an energy level set blocking away a small neighborhood of the bottom of the right well, $\partial G_1$ is another appropriately  chosen energy level set with its energy higher than the energy of the rank-one  saddle.  See Figure (\ref{Graph2}) for illustrations. Then, the MFPT$_{\rm 2N}$, ${\tilde m}^\epsilon (q_0,p_0)$, from a point $(q_0,p_0)$ near the bottom of the left well  can be computed by solving the following Dirichlet problem \footnote{Note: for  mathematicians, Eq.\,(\ref{MFETEQ}) is a Dirichlet problem; for scientists and engineers,  a Pontryagin equation, and for this paper, the term, MFPT equation, will also be used in certain situations. }:

\begin{eqnarray}\label{MFETEQ}
{\tilde L}^\epsilon {\tilde m}^\epsilon (q,p)&=&-1, \hspace{0.2in} (q,p)\in G\nonumber\\
{\tilde m}^\epsilon (q,p)&=&\,\,\,\,0,  \hspace{0.2in} (q,p)\in \partial G
\end{eqnarray}
where ${\tilde L}^\epsilon$ is the differential operator corresponding  to the kinetic Langevin system (Eq.\,(\ref{EOM_PR_dimensionless}) with ${\mathcal E}=0)$.  And the metastable rate is just the inverse of ${\tilde m}^\epsilon(q_0,p_0)$.

However, this partial differential equation is extremely difficult,  or perhaps impossible to solve, even numerically. Fortunately, Freidlin-Wentzell  Theory of Random Perturbations of Hamiltonian Systems allows us to reduce this difficult problem to something more manageable, namely, by solving a corresponding Dirichlet problem, a linear ODE,  on a subset $\Gamma _G\, (=Y(G))$  of a graph $\Gamma$.  See Eq.\,(\ref{MFET1D}) in Section \ref{1D}.

\subsection{MFPT for a Stochastic Averaged 1D Energy Diffusion on a Graph}

Notice that  this kinetic Langevin system is underdamped (see Eq. (\ref{under-damped})), has  a Hamiltonian $H(q,p)$ that is its only first integral, and has two stable equilibria and a rank-one saddle.   Its long time behavior can be described by a diffusion process on a graph. Inside each edge, the process is defined by the standard average procedure, but for the whole process, a gluing condition is needed at an interior vertex corresponding to the energy level set that contains the rank-one saddle. The differential operator of this energy process can be used to set up a Dirichlet problem on the graph whose solution can be used to  approximate the MFPT for the original problem.  See references \cite{FrWen12, FrWeb98, FrWeb01, Fr98, StVa97} for details.  However, since these materials scatter in many publications, for the convenience of readers, we will summarize below the most relevant ones,  make them applicable to our problem, and  with similar notations used in the book by Freidlin and Wentzell \cite{FrWen12} \footnote{For example:  we will use $(q,p)$ instead of $(\theta, p_\theta)$; $\epsilon$ instead of $\nu\kappa$, etc. }.

Given the DNA equations without the electric field
\begin{eqnarray*}
\dot {\tilde q}_t ^\epsilon &=&  \nabla _{p} H(\tilde q_t^\epsilon,\tilde p_t^\epsilon) , \\
\dot {\tilde p}_t^\epsilon &=& -\nabla _q H(\tilde q_t^\epsilon,\tilde p_t^\epsilon) -\epsilon \nabla _p H(\tilde q_t^\epsilon,\tilde p_t^\epsilon ) +\sqrt{2\beta^{-1}\epsilon}\, \dot {\tilde W}_t
\end{eqnarray*}
Even though the perturbation is degenerate, the system is kinetic Langevin and hence hypo-elliptic and the theory of H$\ddot o$rmander  is applicable \cite{Br14, Pa14}. As in the case of non-degenerate perturbations, the process $(\tilde q_t^\epsilon,\tilde p_t^\epsilon)$ in $ R^{2N}$ has, roughly speaking, fast and slow components. The fast component corresponds to the motion along the non-perturbed trajectories on a connected component of a $R^{2N-1}$-dimensional energy surface. The slow component corresponds to the motion transversal to the same connected component. Moreover, the fast component can be characterized by the invariant density of the non-perturbed system on the corresponding connected component.  

To study the slow component, we rescale the time: put $q_t^\epsilon={\tilde q}_{t/\epsilon}^\epsilon,  p_t^\epsilon={\tilde p}_{t/\epsilon}^\epsilon$.  Then $(q_t, p_t)$ satisfies equations
\begin{eqnarray*}
\dot q_t^\epsilon &=& \frac{1}{\epsilon} \nabla_p H(q_t^\epsilon,p_t^\epsilon) , \\
\dot p_t^\epsilon &=& -\frac{1}{\epsilon} \nabla_q H(q_t^\epsilon,p_t^\epsilon) -  \nabla _p H(q_t^\epsilon,p_t^\epsilon)  + \sqrt{2\beta^{-1}}\, \dot W_t
\end{eqnarray*}
where $W_t$ is a N-dimensional Wiener process.

\subsubsection{Standard Stochastic Averaging for Diffusion in One Well} 
First, we restrict our attention only to the left well.  Then the slow component of the process can be characterized by how the energy Hamiltonian $H(q_t^\epsilon, p_t^\epsilon)$ changes. Using Ito formula, we have
\[
H(q_t^\epsilon,p_t^\epsilon)-H(q_0,p_0)=\sqrt{2\beta^{-1}} \int _0^t \nabla_p H(q_s^\epsilon,p_s^\epsilon)\cdot dW_s
\]
\[
+\beta^{-1} \int_0^t  \Delta _p H (q_s^\epsilon,p_s^\epsilon)\,ds
-\int_0^t \nabla_p H^*(q_s^\epsilon, p_s^\epsilon) \nabla_p H(q_s^\epsilon, p_s^\epsilon) \, ds
\]
Applying the standard averaging procedure with respect to the invariant density of the fast motion, with the slow component frozen,  the processes $Y_t^\epsilon =H(q_t^\epsilon, p_t^\epsilon)$ converges weakly in the space of continuous function to the diffusion process  $Y_t$, whose operator is given below by
\[
L=\beta^{-1}\,\frac{ u(h)}{v(h)}\, \frac{d^2}{dh^2}+\beta ^{-1}\, \frac{w(h)}{v(h)}\, \frac{d}{dh}
-\frac{u(h)}{v(h)}\, \frac{d}{dh}
\]
where
\begin{eqnarray*}
u(h)&=& \oint _{C(h)} \nabla_p H^*(q, p) \nabla_p H(q, p) \, \frac{dS}{|\nabla H(q,p)|}\\
w(h)&=& \oint_{C(h)} \Delta _p H (q, p)\,
\frac{dS}{|\nabla H(q, p)|}=\frac{d}{dh}\, u(h)\\
v(h)&=&\oint _{C(h)} \frac{dS}{|\nabla H (q, p)|}
\end{eqnarray*}
Here, $\tfrac{{\rm const.}}{|\nabla H(q, p)|}$  is the invariant density with respect to the volume on the energy surface $C(h)=\{(q, p)\in R^{2N}: H(q, p)=h\}$; $dS$ is the surface element on $C(h)$. The point $0$ which corresponds to the stable equilibrium point in the left well is an exterior vertex and hence inaccessible for the process $Y_t$.  For the proof of $w(h)=u'(h)$, see Appendix A.1.
\subsubsection{FW Theory for Diffusion in Space with  Multiple Wells }

For computing the MFPT and the metastable rate, we need to consider the  whole phase space, in order to keep track of all the trajectories that start from the bottom of the left well, diffuse across the saddle, and reach the  neighborhood of the bottom of the right well.   In this case, level sets of the Hamiltonian may have more than one component. Consider now the graph $\Gamma$ homeomorphic to the set of connected components of level sets of $H(q, p)$.  Then, $\Gamma $ will consist 3 edges, $I_1,I_2,I_3$ and 3 vertices $O_1,O_2,O_3$.  Let $Y: R^{2N}\rightarrow \Gamma$ be this mapping.
$Y(q, p)$ is a point of $\Gamma$ corresponding to the connected components of $C(H(q, p))$ containing point $(q, p)\in R^{2N}$,  and $Y(q, p)=(i(q, p), H(q, p))$, where pair $(i,H)$ forms the coordinates on $\Gamma$.
Consider random processes $Y_t^\epsilon =Y(q_t^\epsilon, p_t^\epsilon)$ on $\Gamma$. These processes converge weakly to the diffusion process $Y_t$ on $\Gamma $. To calculate the characteristics of the limiting process, consider, first, the interior $(i,h)$ of the edge $I_i\subset \Gamma$.  Let $C_i(h)$ be the corresponding level set component.  As before, the process $Y_t$ inside $I_i$ is governed by the operator
\[
L_i=\frac{2\beta^{-1}}{v_i(h)}\, \frac{d}{dh}\,\left(\frac{u_i(h)}{2} \frac{d}{dh}\right)
-\frac{u_i(h)}{v_i(h)}\, \frac{d}{dh}
\]
To determine the limiting process for all $t>0$, the behavior of the process at the vertices,
namely, the exterior vertices, $O_3$ of $I_3$ ($O_3\sim I_3$) and $O_2$ of $I_2$ ($O_2\sim I_2$), and the interior vertex $O_1$ which connects 3 edges $I_1,I_2,I_3$ ($I_3\sim O_1, I_2\sim O_1, O_1\sim I_1$), need to be described.  

\subsubsection{MFPT for 1D Energy Diffusion Process on a Graph }\label{1D}

With this differential operator of $Y_t$ in hand, we can use Freidlin-Wentzell  another result that
\begin{equation*}
\lim _{\epsilon \rightarrow 0}\, \epsilon \, {\tilde m^\epsilon (q, p)}=m(i(q, p),H(q, p)) 
\end{equation*}
to estimate the MFPT$_{2N}$, ${\tilde m}^\epsilon (q, p) $, from the solution, $m(i, h)=m_i(h)$, of the following Dirichlet problem on $\Gamma_G$:
\begin{eqnarray}\label{MFET1D}
\frac{2\beta^{-1}}{v_i(h)}\, \frac{d}{dh}\,\left(\frac{u_i(h)}{2} \,\frac{d\,m_i(h)}{dh}\right)\hspace{-0.08in}
&-& \hspace{-0.08in} \frac{u_i(h)}{v_i(h)}\, \frac{d\, m_i(h)}{dh}\,\,= -1 \nonumber\\
m_i (h)&=& 0,  \hspace{0.2in} (i, h)\in \partial \Gamma_G.
\end{eqnarray}
where 
\begin{enumerate}
\item
$m(i, h)$ is continuous on $\Gamma_G$;
\item 
the gluing condition is satisfied at the interior vertex $O_1$, i.e.,
\begin{equation*}
\alpha_3\, \frac{d\,m_3(h)}{dh}+\alpha_2\, \frac{d\,m_2(h)}{dh}=
\alpha _1\, \frac{d\,m_1(h)}{dh}
\end{equation*}
with
\begin{equation*}
\alpha_i=  \oint _{C_i(Y^{-1}(O_1))} \nabla_p H^*(q, p) \nabla_p H(q, p) \, \frac{dS}{|\nabla H(q,p)|};
\end{equation*}
\item
the exterior vertex, $O_3, $ is an entrance and inaccessible.
\end{enumerate}


\subsection{Computation of Drift and Diffusion Coefficients via MC Method  }

\begin{figure}[ht]
\begin{center}
\includegraphics[width=4.7in]{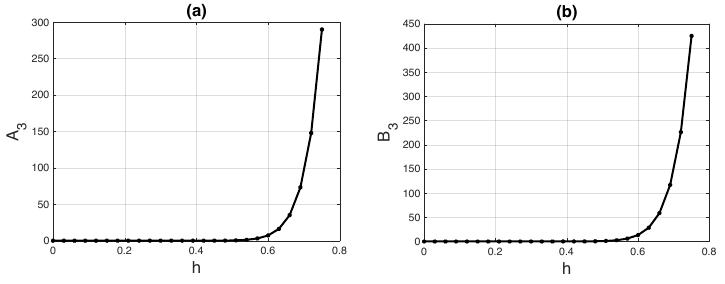}  
\end{center}
\caption{\label{AB} {\footnotesize
To compute the drift and diffusion coefficients for high DOF systems like DNA, we need to reduce the dimension of the volume integrals of $u_i, v_i$ further by integrating  $dp'=dp_2\ldots  dp_N$ part analytically first, and then employing the more sophisticated Metropolis Hastings Monte Carlo method to evaluate the remaining parts $A_i, B_i$ given by Eq.\,(\ref{ABFormula}). 
Figure (a) shows the coefficient 
$A_3(h)$ as a graph of a function of $h$ which is the energy coordinate on $I_3$.  
Similarly Figure(b) shows the graph for the coefficient $B_3(h)$.  Moreover, $A_1(h), A_2(h), B_1(h), B_2(h)$ can also be computed in the same way.
}}
\end{figure}

To solve this Dirichlet problem on $\Gamma_ G$, Eq.\,(\ref{MFET1D}), we need to employ numerical method to find its coefficients. But for high-dimensional system like DNA, we may need to do some simplification first.

(1) We will rewrite it  compactly as
\[
\frac{2\beta^{-1}}{v_i(h)\,\,e^{-\Psi _i(h)}} \, \, \frac{d}{dh}\,\left(e^{-\Psi_i(h)}\,\, \frac{u_i(h)}{2}\,\, \frac{d\,m_i(h)}{dh}\right)
=-1
\]
where $\Psi_i(h)=\int^{h(I_i)} \frac{1}{\beta^{-1}}\,dy$.

(2) Since the surface of $C_i(h)$ can be represented as a $(2N-1)$-dimensional graph in $R^{2N}$, surface integrals for the coefficients, $u_i(h), v_i(h)$, can be transformed into volume integrals  over the domains in the corresponding regions of the phase space \cite{DeZh07a, DeZh07b, CaZh17}.  For a few DOF systems,  coefficients  $u_i, v_i$ can be computed from these volume integral formulas using simply the Monte Carlo method.  But for high DOF systems like DNA, we need to reduce the dimension of the volume integrals further by integrating  $dp'=dp_2\ldots  dp_N$ part analytically first
\cite{DeZh07b}, and then employing the more sophisticated Metropolis Hastings Monte Carlo method \cite{Be07, GaKo05} to evaluate the remaining parts $A_i, B_i$ given below by:
\begin{eqnarray}\label{ABFormula}
A_i(h)&=& \frac{1}{2} \int_{\Sigma_i(h)} (h-V(q))^{\frac{N}{2}}\, dq =u_i(h)/2c \nonumber\\
B_i(h)&=&\int_{\Sigma_i(h)} (h-V(q))^{\frac{N}{2}-1}\, dq=v_i(h)/c
\end{eqnarray}
where $\Sigma_i (h) =\{q: V(q) \leq h\}$.  See Appendix \ref{A2} for more details for the derivation of Eq. (\ref{ABFormula}).  

(3) Figure (\ref{AB}) shows the coefficients,  $A_3(h), B_3(h)$, as a graph of a function of the energy $h$,  computed with the above method. Similarly, $A_1(h), A_2(h), B_1(h)$ an $ B_2(h)$ can also be obtained in the same way, with the same geometric data of our DNA model collected below:
\begin{itemize}
\item
bottom of the left well $(\theta, p_\theta)$ where $\theta _n=-2.6906, \, p_{\theta n}=0, \,\, n=1,\ldots, N$, and with energy $h=0$; bottom of the right well $(\theta, p_\theta)$ where $\theta _n=2.6906, \, p_{\theta n}=0$  and with energy $h=0$;
\item
saddle $(\theta, p_\theta)$ where $\theta _n=0,\, p_{\theta n}=0$, and with energy level $h=0.75$;
\item
$\partial G_1$ is an energy level set with energy $h_1=2.2$ which is much higher than the energy of the saddle;
\item
$\partial G_2$ is an energy level set with energy $h_2=0.15$ which blocks away a small neighborhood of the bottom of the right well;
\item
hence, the energy coordinate $h$ on $I_3$ is from 0 to 0.75; the energy coordinate on $I_2$ is from 0.15 to 0.75; the energy coordinate
on $I_1$ is from 0.75 to 2.2.
\end{itemize}
(4) If we set $M_i=2\beta^{-1}m_i$, then the Dirichlet problem is now given by
\[
\frac{1}{B_i(h)\,\,e^{-\Psi_i(h)}} \, \, \frac{d}{dh}\,\left(e^{-\Psi_i(h)}\,\, A_i(h)\,\, \frac{d\,M_i(h)}{dh}\right)
=-1
\]
\begin{equation}
M_i(h)=0,  \hspace{0.2in} (i,h)\in \partial \Gamma_G
\end{equation}
where 
\begin{itemize}
\item
$M_i(h)$ is continuous on $\Gamma_G$;
\item 
$M_i(h)$ satisfies the gluing condition at the interior vertex $O_1$;
\[
\alpha_3\, \frac{d\,M_3(h)}{dh}+\alpha_2\, \frac{d\,M_2(h)}{dh}=
\alpha _1\, \frac{d\,M_1(h)}{dh}
\]
\[
\alpha_i=  A_i(h)|_{h=h_{O_1}}
\]
\item
the exterior vertex, $O_3, $ is inaccessible.
\end{itemize}
(5) Moreover, the MFET$_{2N}$, $\tilde m^\epsilon (q, p)$, is given by
\[
\lim _{\epsilon \rightarrow 0}\, \epsilon \, {\tilde m^\epsilon (q, p)}=M(i(q, p),H(q, p))/2\beta^{-1} 
\]
and can be estimated by $M(i(q, p),H(q, p))/{2\beta^{-1}\epsilon} $.

\subsection{Solutions for the 1D MFPT Equation on a Graph}

After integrations over segments $I_i$, we obtain
\begin{equation}\label{MFET-Int}
M(i, h)=\int ^{h(I_i)}\left[\int ^z -B_i(y)\,\,e^{-\Psi_i(y)}\,dy \right]\frac{2e^{\Psi_i (z)}}{A_i(z)}\, dz
+b_i \int ^{h(I_i)} \frac{2e^{\Psi_i (z)}}{A_i(z)}\, dz +c_i
\end{equation}
with six constants of integration $b_i, c_i$.   They can be determined uniquely by the following six conditions:
\begin{enumerate}
\item
two boundary conditions
at $h_1,h_2$,
\item
two continuous conditions at the interior vertex $O_1$, 
\item
one gluing condition at the interior vertex $O_1$, and 
\item
one exterior vertex condition at $O_3$. 
\end{enumerate}
And their values are given below:  
\begin{eqnarray*}
b_1\hspace{-0.1in} &=& \hspace{-0.1in}18.8310326781; \hspace{0.44in}
b_2 =  0.4189904781; \hspace{0.1in}
b_3 = 0\\
c_1 \hspace{-0.1in}&=&\hspace{-0.1in}  1.0944699430\times10^{15};
\hspace{0.1in} c_2 = 0;  \hspace{0.83in}
c_3 = 1.09446994586\times10^{15}
\end{eqnarray*}
See Appendix \ref{A3} for more details.

Moreover, if we denote the values of the first and second integration terms of Eq.\,(\ref{MFET-Int}) by $-[[B]]^{h(I_i)}$  and $[A]^{h(I_i)}$, respectively,  then the MFPT$_{2N}$, $\tilde m^\epsilon (q_0,p_0)$, that starts from a point $(q_0,p_0) $ near the bottom of the left well and exits at the boundary $\partial _2$ of the left well  can be approximated by $M_3(h(O_3)^+)/(2\beta^{-1}\nu \kappa)$ where
\begin{eqnarray}
M_3^\epsilon (h(O_3)^+)/(2\beta^{-1}\nu\kappa)&=&(-[[B_3]]_{h(O_3)^+}^{h(O_3)^+}+c_3)/(2\beta^{-1} \nu \kappa)
=c_3/(2\beta^{-1}\nu \kappa)\nonumber \\
&=&  6.244878692881616\times 10^{18}
\end{eqnarray}

Here, we need to choose the value of $h_1$ large enough so that most of the trajectories  do not exist from $\partial G_1$.  See the comments in Section \ref{Observations}.

\begin{figure}[ht]
\begin{center}
\includegraphics[width=4.7in]{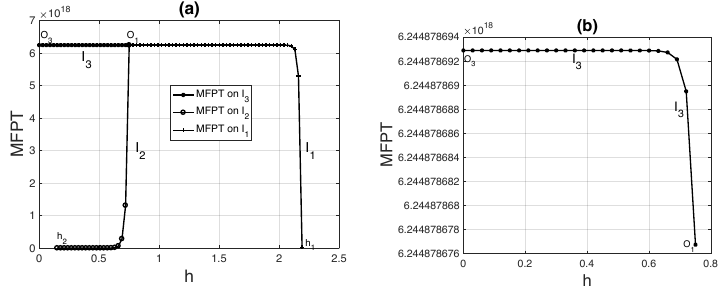} 
\end{center}
\caption{\label{MFET} {\footnotesize
(a) The graph shows the MFPT from the bottom of the left well near $O_3$ (at $C_3(h_0)$, with $h_0$
slightly larger than $0$), to first exit near the bottom of the right well at $C_2(h_2)$, with $h_2=0.15$, or first exit the region $G$ above the saddle at $C_1(h_1)$, with $h_1=2.2$.  
(b) The graph is an enlargement of the first segment of
Figure (\ref{MFET})(a).}}
\end{figure}

\subsection{Results and Comments}

\subsubsection{The Mesoscopic DNA is Robust Against Noise}

Hence, the mean life span of collective base-flipping of the mesoscopic DNA, $T_{DNA}$, in the natural time should be
\[
6.244878692881616\times 10^{18}\times 0.9846 \,\, {\rm ps}=6.1487\times 10^{18}\,\, {\rm ps}
\]
which is about 10 weeks.  Since the life time of a skin cell is about 2 to 4 weeks, and $T_{DNA}>T_{Cell}$,  we can reasonably conclude that it is robust against noise and will survive in a normal situation without radiation.

\subsubsection{FW Theory is Essential in  Studying Metastable Rate}\label{Observations}

Figure (\ref{MFET})(a) shows the MFPT from the bottom of the left well at $C_3(h_0)$, with $h_0$ slightly larger than $0$, to first exit near the bottom of the right well at $C_2(h_2)$,  with $h_2=0.15$, or first exit the top of the region $G$ at $C_1(h_1)$, with $h_1=2.2$.  Figure (\ref{MFET})(b) is an enlargement of the first segment of Figure (\ref{MFET})(a). By studying the results encapsulated in Figure  (\ref{MFET}) closely, the diffusion process can be broadly divided into 3 consecutive segments: 
\begin{enumerate}
\item
in the first segment: it takes only less than  $10^{-6}\,\%$ of the total MFPT ($1.6163 \times 10^{10}$  unit of time) for the system to diffuse from the bottom of the left well to first exit the top of the left well at the level set $C_3(h)$ with $h=0.75$, which is the height of the well;
\item
but  in the second segment: it requires an extremely long time, namely, almost $80\,\%$ of the total MFPT ($4.9329\times 10^{18}$ unit of time) for the system to diffuse across the rank-one saddle, to first pass the top of the right well at the level set $C_2(h)$ with $h=0.75$, and to first exit the level set $C_2(h)$ with $h=0.72$, which indicates that the system has just slightly entered into the right well; 
\item
in the last segment:  it takes another additional  $20\,\%$ of the total MFPT for the process to finally first exit near the bottom of the right well at $C_2(h)$ with $h=0.15$.    
\end{enumerate}

From these observations, we would like to make a couple of remarks:
\begin{enumerate}
\item
In studying the MFPT and metastable rate of a diffusion process, many papers use the standard Hamiltonian stochastic averaging method and restrict their attention to only one well. For a few DOF system, the results may be fine.  But for a large DOF system, the results can be way off.  We highly recommend others to use  Freidlin-Wentzell method.  
\item
This method, in our opinion, can be readily extended  to study the meta-stable rates and branching ratios, etc, of other multiple-well Langevin systems with weak noise, as long as the Hamiltonian is the only first integral and it satisfies some other general requirements, see the references \cite{FrWeb01, FrWen12} for more details. 
\end{enumerate}


\section{But It is Vulnerable to  Specific THz Fields }\label{CT}

After showing that the mesoscopic DNA is robust against noise, we would like to quantify the requirements for specific electric fields that can trigger its collective base-flipping dynamics and cause large amplitude separation of base pairs. Here, we will use the newly developed theory of resonant enhancement of the rate of metastable transition \cite{ChTa22}, which has built on the work of large deviation theory \cite{FrWen12} and the results of researches in physics \cite{DyRa97, SmDy97, DyGo01, AsKa08, Ri72, SoTa19}.

\begin{figure}[ht]\begin{center}
\includegraphics[width=4.7in]{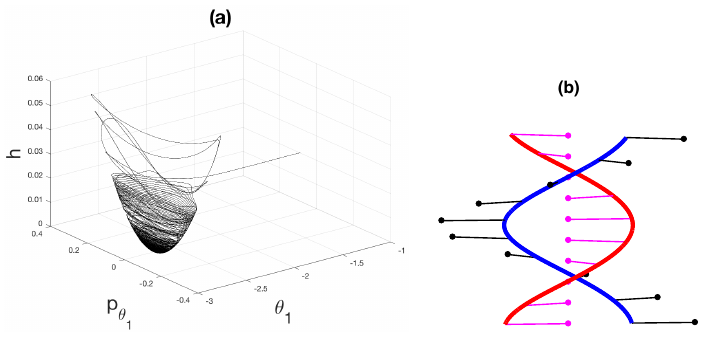}
\end{center}
\caption{\label{HC} {\footnotesize   To apply the theory of resonant enhancement, we need to 
numerically compute the most probable path (MPP) which is essentially the heteroclinic connection between the metastable equilibrium point at the bottom of the left well and  the rank-one saddle. 
See the text for more details.
Figure (a) shows the 3D energy contour plot of the projection of the numerically computed $2N$-dimensional MPP onto 
the   $(\theta_1,p_{\theta_1})$ phase plane. The other similarly drawn 3D energy contour plots on
the $(\theta_n, p_{\theta_n})$ phase planes all share the similar feature.  The MPP allows us to study the resonant enhancement 
of the base-flipping dynamics across the saddle whose configuration in the physical space is shown
in Figure (b). }}
\end{figure}

Below, we will first briefly summarize the main result of this new theory of resonant enhancement and then apply it to our DNA model. We will find the requirements for the specific electric fields in the following steps: (1) we will obtain numerically the most probable path (MPP) which is essentially the heteroclinic connection between the metastable equilibrium point at the bottom of the left well and  the rank-one saddle;  (2) we will derive analytically the formula for the work done by the electric field along the MPP that winds from the bottom to the top of the potential well; (3) we will use this formula to find, via numerical methods, the requirements for the specific electric  fields  and conclude that the mesoscopic DNA is vulnerable to THz fields with frequencies around 0.5 THz and with amplitudes in the order of 450 kV/cm. At the end of the section, (4) we will show that our semi-analytical results compare well with the experimental data of Titova et al. \cite{TiAy13a, TiAy13b, TiAy13c} which have demonstrated that they could affect the function of DNA in human skin tissues by THz pulses with frequencies around 0.5 THz and with a peak electric field at 220 kV/cm;  (5) we will also briefly discuss the need to model the THz pulses and the need to have a better result on the issue of pre-factor.

\subsection{The Theory of Resonant Enhancement of Metastable Rate}

For the convenience of readers, we will restate below the main result of Chao and Tao \cite{ChTa22}, namely, Theorem 2.7,  in the similar notations as in the original paper.\\

\noindent
{\bf Theorem 2.7} Consider $n$ particles in $R^d$ whose equation of motion is given by the underdamped kinetic Langevin system perturbed by a periodic force $f(x,t)$: 
\begin{eqnarray}\label{CTEQ}
dx &=& v\,dt \nonumber \\
dv&=&-\Gamma v\,dt -\nabla V(x)\, dt +\epsilon  f(x,t)\,dt +\sqrt{\mu}\, \Gamma ^{\tfrac{1}{2}}\, dW
\end{eqnarray}
Assume that
\begin{enumerate}
\item
$\mu <\epsilon$,
\item
$x_u$ is the only attractor on the separatrix between the basins of attraction of $x_a$ and $x_b$, and the heteroclinic connection  $x_a$ to $x_b$ exists in the noiseless ($\mu=0$) and forceless ($\epsilon=0$) system,
\item
$\epsilon $ is small enough such that  a heteroclinic connection from  $x_a^\epsilon (t)$ to $x_b^\epsilon (t)$ exists in the high order Euler-Lagrange equation. 
\end{enumerate}
Then the transition rate from  $x_a^\epsilon (t)$ to $x_b^\epsilon (t)$ is asymptotically equivalent to $\exp(-S^\epsilon/\mu)$, i.e.,
\begin{eqnarray}\label{DS}
R\asymp e^{-\frac{S^\epsilon}{\mu}} \hspace{0.2in}
{\rm where} \hspace{0.2in}
S^\epsilon &=&2[V(x_u)-V(x_a)]+\epsilon\, \Delta S_e+O(\epsilon^2)\nonumber\\
\Delta S_e = \min _{t_0}\, \Delta S(t_0),\hspace{0.1in} \Delta S(t_0)\hspace{-0.08in}&=&\hspace{-0.08in}-2\int _{-\infty}^{+\infty}
\dot x_h^T(t-t_0)f(x_h(t-t_0),t)\, dt
\end{eqnarray}
and 
$x_h(t)$ satisfies
\begin{equation}\label{UpHillEQ}
\ddot x_h -\Gamma \dot x_h+\nabla V(x_h)=0, \hspace{0.3in} x_h(-\infty)=x_a, \;\;\; x_h(+\infty)=x_u
\end{equation}

\subsection{Its Application to the New DNA Model} 

Recall  here the non-dimensional equations of motion of our DNA model with noise and radiation 
\begin{eqnarray*}
d\theta _n &=& p_{\theta _n} \, dt \\
dp_{\theta{n}} &=& -\nu \kappa  \, p_{\theta _n}\, dt -((\theta _{n+1}-2\theta _n+\theta _{n-1})-
\kappa U'(\theta_n))\, dt\\
&& +{\mathcal E}\, (\kappa \sin(\theta _n+(n-1)\pi/5)\cos \Omega t) \, dt+\sqrt{2\beta ^{-1}}( \nu \kappa )^{\tfrac{1}{2}}\, dW
\end{eqnarray*}
which are almost identical to the Equations (\ref{CTEQ}), with
\[
V(\theta)=\sum_{n=1}^N\left[\frac{1}{2}(\theta_n-\theta_{n-1})^2+\kappa 
\left(e^{-a(1+\cos\theta _n-d_0)}-1\right)^2\right]
\]
\begin{equation}\label{f}
f(\theta,t)=\kappa \sin(\theta _n+(n-1)\pi/5)\cos \Omega t 
\end{equation}
As stated earlier, the parameters are chosen to best represent the typical values for the
opening and closing dynamics of DNA and are given by  
\begin{enumerate}
\item
$\nu \kappa=0.006564$ where $\nu\kappa$ is the friction coefficient, 
\item
$\kappa=\frac{D}{S}=0.025$  where $\kappa$ is the strength of the Morse potential,
\item
$\mu=2\beta ^{-1}=2\frac{k_B T}{S}=0.0267$ where $2\beta^{-1}$  is the strength of the noise.
\end{enumerate}
Moreover, $\mathcal E$ is the moment of force (w.r.t the strength of Morse potential $\kappa$) induced by the electric field in the non-dimensional system. The theory of resonant enhancement requires that ${\mathcal E}>\mu$, which means that $\mathcal E $ needs to be larger than 0.0267.

Below, we will  first, numerically,  find the most probable path (MPP) which is essentially the heteroclinic connection between the metastable equilibrium point at the bottom of the left well and  the rank-one saddle. 

\subsubsection{Computation for the Most Probable Path--Heteroclinic Connection} 

As we have studied in Section \ref{FW}, the DNA model has 2 stable equilibria $W _L, W_R$ and a rank-one saddle $S$ given by  
\begin{eqnarray*}
W _L&=&(-\theta_e,\cdots, -\theta_e; 0,\cdots, 0), \hspace{0.3in}
S=(0,\cdots,0; 0, \cdots, 0),\\
W _R&=&(\theta_e,\cdots, \theta_e; 0,\cdots, 0),
\end{eqnarray*}
where $\theta_e=\arccos (d_0-1)=2.6906 $.  

According to Theorem 2.7, we first need to find the most probable path, i.e, to approximate the heteroclinic connection from  $\theta_L$ to $\theta_S$  of  this noiseless and forceless system by numerically solving the uphill equations,  Eqs.\,(\ref{UpHillEQ}). However, since this is a second order boundary value problem with the  boundary points at $t=\pm \infty$, it poses numerical difficulties.   And the resonant enhancement theory \cite{ChTa22} suggests to tackle them in the following steps:  
\begin{enumerate}
\item
First, reverse the time and solve the corresponding downhill equations from the saddle $S$ to $W _L$ with the  sign flip on the velocity in the uphill equations. 
\item
Then, make the approximation by choosing an initial point as close as numerically feasible to the saddle,  and in the unstable direction of the downhill  vector field linearized at the saddle.
\item
Finally, simulate this initial value problem using fourth-order Runge-Kutta for long enough time, with sufficiently small time steps, and collect the result backward in time.
\end{enumerate}
Figure\,(\ref{HC}) shows the 3D energy contour plot of the projection of this $2N$-dimensional MPP onto the $(\theta_1,p_{\theta_1})$ phase plane. The other similarly constructed 3D energy contour
plots all share the similar feature.

\subsubsection{Work Done Along the MPP and the Rate Enhancement} 

Next, we need to apply the general formula Eq.\,(\ref{DS}) to our DNA model. Following similar steps as in Chao and Tao \cite{ChTa22}, we can show that 
\begin{equation}\label{DS2}
 \Delta S_e
 = -2 \left| \int _{-\infty}^{+\infty} \left[ \sum _{j=1}^{N} 
\kappa \dot x_h^j(t)\, \sin(x_h^j (t)+(j-1)\alpha) \right]\, e^{i\Omega t} \, 
dt\right|=-2\kappa |{\mathcal I(\Omega)}|
\end{equation}
See Appendix B for details.  Here, it is interesting to note that  the integral term in Eq.\,(\ref{DS}),
after rewritten as 
\[
\int _{-\infty}^\infty \,f^T(x_h(t-t_0),t)\,dx_h
\]
can be seen as the work done, by the force $f$,  along the MPP in the configuration space ($x_h$),  and in overcoming the potential difference of the well \cite{SmMc99}
\[\delta V=[V(x_u)-V(x_a)].\]

\begin{figure}[ht]
\begin{center}
\includegraphics[width=4.7in]{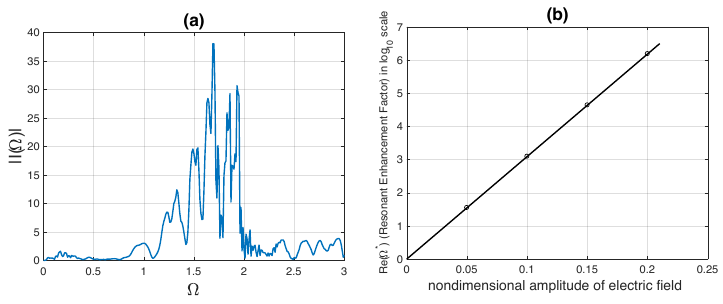} 
\end{center}
\caption{\label{RE} {\footnotesize
(a) The graph shows the relationship between $|{\mathcal I}(\Omega)|$ and $\Omega$.  We observe that there exists a special $\Omega^*=1.7$, called resonant frequency,  at which $|{\mathcal I}(\Omega)|$ peaks. The graph has a broadband character between 1 to 3 (with peak at 1.7).   This is due to the fact that (i) the system has 30 DOF with internal frequencies between 1.5 to 3, and (ii) it is a kinetic Langevin system with weak damping and noise. 
(b) The graph shows the relation between the resonant enhancement factor $R_e(\Omega^*)$ and  the non-dimensional amplitude of the electric field $\mathcal E$. For example, for ${\mathcal E}=0.1$,  we have $R_e(\Omega^*)=1.2411\times 10^3$; and for ${\mathcal E}=0.2$, we have $R_e(\Omega^*)=1.5404\times 10^6$
}}
\end{figure}

Then the escape rate $R$ from $x_L^{\mathcal E} (t)$ to $x_R^{\mathcal E} (t)$ is asymptotically equivalent to $\exp(-S^{\mathcal E} /\mu)$ with $\mu=2\beta^{-1}$, i.e.,
\[
R\asymp e^{-\frac{S^{\mathcal E}}{\mu}} 
\]
where
\begin{eqnarray*}
S^{\mathcal E} &=& 2\,\delta V+{\mathcal E} \, \Delta S_e+O({\mathcal E}^2)\\
&=& 2\,\delta V-2 {\mathcal E}\kappa |{\mathcal I}(\Omega)|+O({\mathcal E}^2)
\end{eqnarray*}
It can be rewritten as 
\begin{equation}\label{R}
R=C(\mu)\times e^{-\frac{S^{\mathcal E}}{\mu}} =C(\mu)\times e^{-\frac{2\,\delta V}{\mu}}\times
e^{\frac{2{\mathcal E}\kappa |{\mathcal I}(\Omega)|}{\mu}}\times e^{\frac{-O({\mathcal E}^2)}{\mu}}
\end{equation}
where the first exponent factor  is related to the metastable rate of the forceless system, the second exponent factor $R_e$ is related to the resonant enhancement due to the electric field  with angular frequency $\Omega$,  and the first factor $C(\mu)$ is called prefactor whose appearance is due to fact that the result of Theorem 2.7 is only valid up to asymptotic equivalence \footnote{See Section \ref{prefactor} for further discussions}.

Now, with $(x_h(t), \dot x_h(t))=(\theta (t), p_{\theta}(t))$ of the heteroclinic orbit and the formula given by Eq.\,(\ref{DS2}) in hand, we can study the dependence of $\Delta S_e$ on the input frequency $\Omega$ via ${\mathcal I}(\Omega)$. For each $\Omega$, we compute $|{\mathcal I}(\Omega)|$ by numerically approximating the integral via piecewise trapezoidal quadrature with high enough resolution. Figure (\ref{RE})(a) shows the relationship between $|{\mathcal I}(\Omega)|$ and $\Omega$.  We observe that there exists a special $\Omega^*=1.7$ at which $|{\mathcal I}(\Omega)|$ peaks. In the theory of resonant enhancement, this frequency $\Omega^*$ is called the resonant frequency. It is interesting to note here that the graph has a broadband character between 1 to 3 with peak at 1.7.   This is due to the fact that (i) the system has 30 DOF with internal frequencies between 1.5 to 3, and (ii) it is a kinetic Langevin system with weak noise. 
   
Moreover, if we call 
\begin{equation}\label{Epart}
R_e(\Omega)=e^{\frac{2{\mathcal E}\kappa |{\mathcal I}(\Omega)|}{\mu}}
\end{equation} 
loosely as the rate enhancement factor, then $R_e({\Omega^*})$ at the resonant frequency  $\Omega^*$ can be seen as the resonant enhancement factor.  Figure (\ref{RE})(b) shows the relation between the resonant enhancement factor $R_e(\Omega^*)$ and  the non-dimensional amplitude of the electric field  ${\mathcal E}$ (w.r.t. the strength of the Morse potential).  For example, for ${\mathcal E}=0.1$,  we have $R_e(\Omega^*)=1.2411\times 10^3$; and for ${\mathcal E}=0.2$, we have $R_e(\Omega^*)=1.5404\times 10^6$

\subsubsection{DNA is Vulnerable to THz Fields with High Amplitudes} 

With the resonant enhancement relationship encapsulated in Figure (\ref{RE})(b) in hand, we are ready to find the requirements for  specific electric fields that can trigger its collective base-flipping dynamics and cause large amplitude separation of base pairs. 

\begin{enumerate}
\item
First,  we would like to recall our semi-analytical result on the mean first passage time (MFPT) of our forceless system.  In Section \ref{FW}, we have estimated that the life span of our mesoscopic DNA is given by
\[
T_{DNA}\approx 6.2449*10^{18}\,\,{\rm ps}=6.2449\times 10^6\,\, {\rm s}
\]
which is about 10 weeks. Correspondingly, its metastable rate is given by
\[
R_0 \approx 1.601 \times 10^{-7}/{\rm s}
\]
Hence, for $\mathcal E$ between 0.1 and 0.2, with its metastable rate enhancement factor between $1.2411\times 10^3$ and $1.5404\times 10^6$, it seems reasonable to claim that electric fields with the resonant frequency $\Omega^*=1.7$ and with $\mathcal E$ between 0.1 and 0.2  will  induce collective base-flipping and cause large amplitude separation of base pairs \footnote{See Section \ref{prefactor}  for further discussions}. 
\item
Before comparing the above results with the experimental data, we need to recall a  couple of formulas from Section \ref{Formula} that link $\Omega$ and $\mathcal E$ of the non-dimensional system to the frequency $\nu_0$ and amplitude $E$ of electric fields.   According to the formula, Eq.\,(\ref{Omega}), for  $\Omega^*=1.7$, the resonant frequency $\nu_0^*$ of the required electric field is  given by
\[
 \nu_0^*=1.0156\times(1.7/2\pi)=
0.1616 \times 1.7\,\, {\rm THz}=0.2748\,\, {\rm THz}
\]
which is a THz field that peaked at 0.2748 THz with a bandwidth 0.16-0.48 THz.  This semi-analytical result is consistent with the experimental data of Titova et al. The broadband THz pulses in their experiments had an amplitude spectrum that peaked at 0.5 THz, with a bandwidth of 0.1-2.0 THz.   
\footnote{See Section \ref{Titova} for more details.} 
\item
According to the formula Eq.\,(\ref{E}), for $\mathcal E$ between 0.1 to 0.2, amplitudes of electric fields $E$ are given by
\begin{equation*}
E=4.5033\times 10^8\times {\mathcal E}=4.5033\times {\mathcal E}\,\, {\rm (MV/cm)}
\end{equation*}
i.e., between $453.33 \,\, {\rm (kV/cm)}$ and  $906.66\,\, {\rm (kV/cm)}$. While these are indeed electric fields with high amplitude, they are also within the same order as the experimental data in Titova et al.  whose peak electric field is equal to 220 kV/cm. \footnote{See Section \ref{Titova} for more details.} 
\end{enumerate}

In conclusion, we have shown that  DNA is vulnerable to THz fields with high amplitudes and our semi-analytical result is consistent with the experimental data of Titova et al.  In Section \ref{Titova}, we will elaborate further on the experimental setup of Titova et al. and discuss the issues of non-thermal effects, pulses vs waves, etc.

\subsubsection{The Need to Have a Better Result on the Issue of Prefactor}\label{prefactor}

Recall that the proof in Chao and Tao \cite{ChTa22} is based on the assumption that $\mu <\epsilon $ so that it can first uses the  large deviation principle and then the asymptotic analysis for the maximum likelihood path.  Hence the result is in the form of asymptotic equivalence, i.e., the transition rate $R$ is given by
\[
R \asymp \exp{((-2\,\delta V-\epsilon \Delta S_e+{\mathcal O}(\epsilon^2))/\mu)}
\]
which can be rewritten as
\[
R=C(\mu)\times \exp{(-2\,\delta V)/\mu}\times \exp{((-\epsilon \Delta S_e+{\mathcal O}(\epsilon^2))/\mu)}
\]
with $C(\mu)$ as a pre-factor.

If $\epsilon$ and $\mu$ is not related, it looks tempting to let $\epsilon \downarrow 0$ and one will have
\[
R_0 
=C(\mu)\times \exp{(-2\,\delta V)/\mu}
\] 
and then
\[
R=R_0\times \exp{((-\epsilon\delta S_e+{\mathcal O}(\epsilon^2))/\mu)}
\]
This new formula, if true,  will decompose the transition rate $R$ into the product of two parts, the first part is the metastable rate of the forceless system, and the second part is the enhancement rate due to the forcing.  Since the first part can be obtained via Freidlin-Wentzell  theory of Hamiltonian Perturbations and  the second part can be obtained via theory of metastable rate enhancement, we will be able to circumvent the issue of pre-factor.  

However, since the proof in Chao and Tao \cite{ChTa22} is based on $\mu<\epsilon$, the above reasonings need to be scrutinized carefully.  And more work is needed to obtain a sharper result.
\subsection{Comparison with Experiments of Titova et al.}\label{Titova}

In a series of papers \cite{TiAy13a, TiAy13b, TiAy13c}, Titova et al. analyzed the effects of intense, broadband, picosecond-duration THz pulses on the DNA integrity in an artificial human skin tissue. Their work yielded an experimental demonstration of the disruptions of DNA function in an exposed skin tissue. Below, we will give a brief description of their experimental setup, and summarize two of their main findings that are relevant to our study, namely, (i) the effect is non-thermal, and (ii) the high peak electric field is  likely responsible for the disruptions in DNA dynamics, or even DNA damage. 

\subsubsection{The Experimental Setup of Titova et al.}

The skin tissue was exposed to THz pulses with 1 kHz repetition rate.  Each pulse had  a pulse energy up to $1\, \mu $J, and an amplitude spectrum that peaked at 0.5 THz with a bandwidth of 0.1-2.0 THz. The time of exposure was 10 minutes long. Duration of a THz pulse was 1.7 picosecond.  Spot size of THz beam at the focus was 1.5 mm in diameter. Now, let $W_{\rm THz}$ be the pulse energy, $\tau$ be the pulse duration, $A$  be  the area of beam focus with radius $r$ equal to 0.75 mm, and $Z_0=377$ ohm  be the impedance of the free space.  Titova et al. were able to use these experimental data and  a couple of formulas to make their points.

\subsubsection{The Effect was  Non-Thermal in Their Experiments}

As expected, by using picosecond duration THz pulse source with low repetition rates, the THz-pulse-induced heating effect was negligible. The exposure was carried out at 21$^\circ$ C and the time-averaged THz power density, which could be computed according to the following formula
\[
(W_{THz}/A)\times 10^3/s =57\,\,{\rm mW/cm^2},
\]
was quite low.  Applying the thermal model in Kristensen et al. \cite{KrWi10}, the estimated temperature increase at beam center due to THz exposure was less than 0.7$^\circ$ C. Moreover, none of the heat shock protein encoding genes were differentially expressed in the THz-exposed tissues. These allowed them to conclude that the biological effects of picosecond duration THz pulses were non-thermal as average power levels were insufficient to cause significant heating, and the levels of temperature-change-sensitive genes and proteins were unchanged.

\subsubsection{The High Peak Electric Field Might Cause the Disruptions of DNA Dynamics } 

However, THz pulses with picosecond duration and 1 $\mu $J energy have high instantaneous electric field which might cause disruptions in DNA dynamics and even DNA damage. For a THz pulse  with Gaussian spatial profile, its peak electric field could be calculated from the pulse energy as follow:
\[
E_{\rm THz}^{\rm Peak}=\sqrt{\frac{4 Z_0W_{\rm THz}}{\tau A}}
\] 
As it turned out, the peak electric field was 220 kV/cm which was extremely high.  They concluded that these high peak electric fields were likely responsible for the observed THz-pulse-driven cellular effects.

Moreover, they mentioned in their papers the work of Alexandrov et al. in a favorable light and hoped to understand better the mechanism, and the specific requirements for the intensity and frequency of the radiation.
\subsubsection{Our Work May Have Suggested the Mechanism and Provided the Requirements}

As mentioned in the {\bf Introduction}, by building on  our earlier work \cite{KoOw13}, the insights from the controversy around the paper of Alexandrov et al.,   and especially the non-thermal experimental results of Titova et al., we develop a  
new mesoscopic DNA model. 
We start with the DNA in the fluid with thermal noise. But we disregard the additional thermal effects by ignoring radiation-fluid interactions, 
in order to  focus our attention mainly on the effects of radiation on the DNA via coupling the electric field  with the DNA bases.
These new mesoscopic DNA model allows us to employ the Freidlin-Wentzell theory of stochastic averaging and the newly developed theory of resonant  enhancement.

Our study has shown that 
while the mesocopic DNA is metastable and robust to environmental effects, 
it is vulnerable  to certain frequencies that could be targeted by  specific THz fields for triggering
its collective base-flipping dynamics and causing large amplitude separation of base pairs.  
Moveover,
our semi-analytic  estimates show that the required  fields should be THz fields with frequencies around 0.28 THz and with  amplitudes in the order of 450 kV/cm. 
These estimates compare well with the experiments of Titova et al.,  which have demonstrated that they could affect the function of DNA in human skin tissues by THz pulses with frequencies around 0.5 THz and  with a peak electric field at 220 kV/cm.   

We believe that our results may have suggested resonance as a possible mechanism for
the experiments of  Titova et al. and  may have provided the specific requirements for the intensity and frequency of the radiation.

\subsubsection{The Need to Model THz Pulses}\label{pulse}

But of course more work is still needed to take into consideration the issue of modeling the pulses.  In the experiments of Tiotva et. al., they skillfully used intense THz pulses with low repetition rates to obtain  a sufficiently high peak electric field and to disrupt  the DNA dynamics, or even cause DNA damage, in a non-thermal setting. In our present study, we try to theoretically investigate  the mechanisms and requirements for such specific electric fields using the existing tools in the theory of control of stochastic mechanical systems. By ignoring the additional radiation-fluid interaction, we get ourselves into the non-thermal setting, minimize the impacts of the issue of waves vs pulses, and enable us to obtain some interesting and important results.   And we wish that in the near future,  further researches will be able to  extend the present theory of resonant enhancement  to the case where the driving  radiation are pulses. 


\section{Conclusion and Future Work}
Our new mesoscopic DNA model is a torsional model that takes into account  not only the issues of helicity and the coupling of an electric field with the base dipole moments, but also includes  environmental effects such as fluid viscosity and thermal noise.  And all the parameter values are chosen to best represent the typical values for the opening and closing dynamics of a DNA. While it is different from the one used by Alexandrov et al., our semi-analytical results do suggest  similar conclusion: DNA natural dynamic  can be significantly affected by specific THz radiation exposure.

This new mesoscopic DNA model allows us to employ Freidlin-Wentzell theory of stochastic averaging and the newly developed theory of resonant  enhancement.  
Our semi-analytic  estimates show that the required  fields should be THz fields with frequencies around 0.28 THz and with  amplitudes in the order of 450 kV/cm. 
These estimates compare well with the experiments of Titova et al.,  which have demonstrated that they could affect the function of DNA in human skin tissues by THz pulses with frequencies around 0.5 THz and  with a peak electric field at 220 kV/cm.  
And we believe that our results may have suggested resonance as a possible mechanism for the experimental results of  Titova et al. and  may have provided the specific requirements for the intensity and frequency of the radiation.

But more work is needed to take into consideration the issue of modeling the pulses.   
And we wish that future research will extend the present theory of resonant enhancement  to the case where the radiation are pulses, as suggested in Section \ref{pulse}.  Besides, a better result on the issue of prefactor mentioned in Section \ref{prefactor} will also be important.  As for the issue of inhomogeneity, our initial explorations and computations have convinced us  that the frequencies and amplitudes of the required THz fields are definitely sequence dependent (A,T,G,C), but we will leave them to our next paper. 

However, even as it stands right now, the new DNA model and the semi-analytical results may have already provided a better understanding of the resonant mechanism and hopefully will inspire new experimental designs that may settle this controversy.
Moreover, we believe our semi-analytical methods may be useful in studying the  metastable rate and its enhancement  for  chemical as well as other bio-molecular systems.


\subsection*{Acknowledgments}
We would like to dedicate this paper to the memory of Professor R. Stephen Berry for his generosity,
helpful suggestions, and consistent encouragement. 
HO acknowledges support the Air Force Office of Scientific Research under MURI award number FA9550-20-1-0358 (Machine Learning and Physics-Based Modeling and Simulation).
MT is partially supported by NSF
DMS-1847802, NSF ECCS-1942523, Cullen-Peck Scholarship, and Emory-GT AI.Humanity Award.

\subsection*{AUTHOR  DECLARATIONS}

\subsection*{Conflict of Interest}
The authors have no conflicts to disclose.

\subsection*{Author Contributions}
{\bf W. S. Koon}: Conceptualization (equal); Formal analysis (lead); Methodology (lead);  
Software (equal); Writing-original draft (lead); Writing-review \& editing (equal). {\bf H. Owhadi}: Conceptualization (equal); Funding acquisition (equal); Methodology (supporting); Writing-review \& editing (equal). {\bf M. Tao}: Conceptualization (equal); Funding acquisition (equal); Methodology (supporting); Software (equal); Writing-review \& editing (equal).  {\bf T. Yanao}: Conceptualization (equal); Methodology (supporting); Writing-review \& editing (equal).

\subsection*{DATA  AVAILABILITY}

The data that support the findings of this study are available from the corresponding author upon reasonable request.

\section*{APPENDIX}

\appendix

\section{FW Theory for Diffusion in Space with Multiple Wells}

\subsection{A Formula in Stochastic Averaging}\label{A1}

Using the formula 
\[
\frac{d}{dh}\oint _{C(h)} G\cdot \frac{\nabla H}{|\nabla H|}\, dS=
\oint_{C(h)} \nabla \cdot G\, \frac{dS}{|\nabla H|}
\]
we can show that  
\begin{eqnarray*}
\frac{d}{dh} \, u(h)&=& \frac{d}{dh}\, \oint _{C(h)}\, (0,\nabla _p H)\cdot \frac{\nabla H}{|\nabla H|}\,
dS \nonumber\\
&=&\oint _{C(h)} \, \nabla \cdot (0,\nabla _p H) \,\frac{dS}{|\nabla H|}=\oint_{C(h)}
\Delta _pH\, \frac{dS}{|\nabla H|}=\, w(h)
\end{eqnarray*}

\subsection{Computation of Drift and Diffusion Coefficients }\label{A2}

Since the surface of $C_i(h)$ can be represented as a $(2N-1)$-dimensional graph in $R^{2N}$,
\[
p_1=\pm \sqrt{2(h-V(q))-\sum_{j=2}^{N}\, p_j^2}
\]
surface integrals, $u_i(h)$ and $ v_i(h)$, can be transformed into volume integrals  
over the domains in the corresponding regions of the phase space
\[
\Omega _i(h)=\left\{
(q, p_2,\cdots, p_N): \sum_{j=2}^N \, p_j^2+2V(q)\leq 2h
\right\}
\]
For $v_i(h)$, we have
\[
v_i(h)=\oint_{C_i(h)}\, \frac{1}{|\nabla H|}\, dS=\int_{\Omega _i(h)}
\frac{1}{|\nabla H|}\cdot \frac{|{\nabla H}|}{{\nabla H}\cdot {\bf e}_1}\, dq\,dp'
=\int_{\Omega_i(h)}\, \frac{1}{p_1}\, dq\,dp'
\]
where $dp'=dp_2\cdots dp_N$ 
Similarly, for $u_i(h)$ (and $\alpha _i$)
\[
u_i(h)=\int_{\Omega_i(h)}\, \frac{\sum _{i=1}^N \, p_i^2} {p_1}\, dq\,dp'
\]
For a few DOF systems,  coefficients  $u_i(h), v_i(h)$ can be computed from these
volume integral formulas using simply Monte Carlo method.  But for a high DOF system
like the DNA, we will reduce the dimension further by integrating $dp'$ part 
analytically first and then employing more sophisticate MH Monte Carlo method to evaluate the remaining parts, $A_i(h), B_i(h)$, given below by  
\begin{eqnarray*}
u_i(h)&=&c\int_{\Sigma_i(h)} (h-V(q))^{\frac{N}{2}}\, dq =2cA_i(h)\\
v_i(h)&=&c\int_{\Sigma_i(h)} (h-V(q))^{\frac{N}{2}-1}\, dq=cB_i(h)
\end{eqnarray*}
where $\Sigma_i (h) =\{q: V(q) \leq h\}$.
This is because
\[
v_i (h)=2\int_{\Sigma_i (h)} \int _{B(r(q))}\frac{1}{\sqrt{r(q)^2-\sum_{j=2}^N \, p_j^2}}
\, dp'dq
\]
\[
=c\int_{\Sigma_i(h)} (h-U(q))^{\frac{N}{2}-1}\,dq
\]
where $r(q)=\sqrt{2(h-V(q))}$ is the radius of the ball  $B(r(q))$.
Similar proof works for $u_i(h)$. 

\subsection{Solutions for the 1D MFPT Equation on a Graph}\label{A3}
\begin{itemize} 
\item
After first integration,
\begin{equation*}
\frac{A_i(h)}{2}\,\, \frac{d\,M_i(h)}{dh}
= e^{-\Psi_i(h)}\,\, \left[\int _{h(O_k)}^{h(I_i)} -B_i(y)\,\,e^{\Psi_i(y)}\,dy +b_i\right]
\end{equation*}
For $I_3$  at $O_3$ (exterior), we have $h(I_3)=h(O_3)$.   Since $O_3$ is an exterior vertex,  $A_3(h(O_3))=0$, we have  
\begin{equation*}
b_3=0
\end{equation*}
\item
At saddle $O_1$, the gluing condition allows us to equate the sum of the right hand side for $I_3$
and $I_2$ with those for $I_1$. After simplification, we obtain
\begin{eqnarray*}
b_1&=&e^{\Psi_3(h(O_1))}\left[\int _{h(O_3)}^{h(O_1)} -B_3(y)\,\,e^{-\Psi_3(y)}\,dy \right] \nonumber\\ 
&& +e^{\Psi_2(h(O_1))}\left[\int _{h_2}^{h(O_1)} -B_2(y)\,\,e^{-\Psi_2(y)}\,dy +b_2\right]
\end{eqnarray*}
\end{itemize}

\begin{itemize} 
\item
After further integrations, we impose the continuity condition for $M(i,h)$ at
the interior vertex $O_1$, and after simplifications, we obtain
\begin{equation*}
c_1=c_3+\int_{h(O_3)}^{h(O_1)} \left[\int _{h(O_3)}^z -B_3(y)\,\,e^{-\Psi_3(y)}\,dy \right]\frac{2e^{\Psi_3 (z)}}{A_3(z)}\, dz
\end{equation*}
\begin{eqnarray*}
c_1=c_2&+&\int_{h_2}^{h(O_1)} \left[\int _{h_2}^z -B_2(y)\,\,e^{-\Psi_2(y)}\,dy \right]\frac{2e^{\Psi_2 (z)}}{A_2(z)}\, dz \nonumber \\
&+& b_2\int_{h_2}^{h(O_1)} \frac{2e^{\Psi_2 (z)}}{A_2(z)}\, dz
\end{eqnarray*}
\item
The boundary condition at $h_2$ gives us 
\begin{equation*}
c_2=0
\end{equation*}
\item
The boundary condition at $h_1$ give us
\begin{equation*}
0=c_1+\int_{h(O_1)}^{h_1} \left[\int _{h(O_1)}^z -B_1(y)\,\,e^{-\Psi_1(y)}\,dy \right]\frac{2e^{\Psi_1 (z)}}{A_1(z)}\, dz
+b_1 \int_{h(O_1)}^{h_1} \frac{2e^{\Psi_1 (z)}}{A_1(z)}\, dz
\end{equation*}
\end{itemize}
This system of six linear equations can be solved to provide the values for the parameters
$b_i$ and $c_i$.  After evaluating all the relevant integrals $[[B_i]]^{h(I_i)}$ and 
$[A]^{h(I_i)}$, we obtain
\begin{eqnarray*}
b_1 &=& 18.831032678119339\\
b_2 &=&  0.418990478121480 \\
b_3 &=& 0\\
c_1 &=&  1.094469943027227e+15\\
c_2 &=& 0\\
c_3 &=& 1.094469945860001e+15
\end{eqnarray*}

\section{Theory of Resonant Enhancement of  Metastable Rate }\label{B0}

Following similar steps as in \cite{ChTa22}, we show that 
\begin{equation*}
 \Delta S_e
 = -2 \left| \int _{-\infty}^{+\infty} \left[ \sum _{j=1}^{N} 
\kappa \dot x_h^j(t)\, \sin(x_h^j (t)+(j-1)\alpha) \right]\, e^{i\Omega t} \, dt\right|=-2\kappa |{\mathcal I}(\Omega)|
\end{equation*}
This is because
\begin{eqnarray*}
 \Delta S(t_0)&=& -2\int _{-\infty}^{\infty} \left[\sum _{j=1}^{N} \kappa \dot x_h^j(t) 
 \sin(x_h^j (t)+(j-1)\alpha) \right] \cos(\Omega(t+t_0))\,dt \\
 &=& -e^{i\Omega t_0} \int _{-\infty}^{\infty} \left[\sum _{j=1}^{N} \kappa \dot x_h^j(t) 
 \sin(x_h^j (t)+(j-1)\alpha) \right] e^{i\Omega t}\,dt \\
  && -e^{-i\Omega t_0} \int _{-\infty}^{\infty} \left[\sum _{j=1}^{N} \kappa \dot x_h^j(t) 
 \sin(x_h^j (t)+(j-1)\alpha) \right] e^{-i\Omega t}\,dt \\
 &=&  -2 \cos(\Omega t_0) {\mathcal Re}\left(\int _{-\infty}^{\infty} \left[\sum _{j=1}^{N} \kappa \dot x_h^j(t) 
 \sin(x_h^j (t)+(j-1)\alpha) \right] e^{i\Omega t}\,dt \right) \\
  && + 2 \sin(\Omega t_0) {\mathcal Im}\left(\int _{-\infty}^{\infty} \left[\sum _{j=1}^{N} \kappa \dot x_h^j(t) 
 \sin(x_h^j (t)+(j-1)\alpha) \right] e^{i\Omega t}\,dt \right) \\
&=& -2 \cos(\Omega t_0+\phi) \left| \int _{-\infty}^{+\infty} \left[ \sum _{j=1}^{N} 
\kappa \dot x_h^j(t)\, \sin(x_h^j (t)+(j-1)\alpha) \right]\, e^{i\Omega t} \, dt\right|
\end{eqnarray*}
where
\[
\sin \phi=\frac{{\mathcal Im}\left(\int _{-\infty}^{\infty} \left[\sum _{j=1}^{N} \kappa \dot x_h^j(t) 
 \sin(x_h^j (t)+(j-1)\alpha) \right] e^{i\Omega t}\,dt \right)}
 {\left|\int _{-\infty}^{\infty} \left[\sum _{j=1}^{N} \kappa \dot x_h^j(t) 
 \sin(x_h^j (t)+(j-1)\alpha) \right] e^{i\Omega t}\,dt \right|}
\]
Therefore, 
\[
\Delta S_e={\rm min}\,_{t_0}\, \Delta S(t_0)=-2\kappa |{\mathcal I}(\Omega)|
\]



\begin{thebibliography}{99}

\bibitem{AbEr21}
M. H. Abufadda, A. Erdelyi, E. Pollak, P. S. Nugraha, J. Hebling, J. A. Fulop, and L. Molnar,
Teraherz pulses induce segment renewal via cell proliferation and differentiation overriding the endogenous regeneration program of the earthaorm Eisenia andrei, {\em Biomed. Opt. Express}
{\textbf 2021}, 12, 1947-1961

\bibitem{AlGe10}
B. S. Alexandrov, V. Gelev, A. R. Bishop, A. Usheva, and K.O. Rasmussen, DNA breathing dynamics in the presence  of a terahertz field, {\em Phys. Lett. A} {\textbf 374} (2010) 1214-1217

\bibitem{AsKa08}
M. Assaf, A. Kamenev, and B. Meerson, Population extinction on a time-modulated environment,
{\em Phys. Rev, E}, {\textbf 78} (2008), 041123

\bibitem{Be07}
K. J. Beers, Numerical Methods for Chemical Engineering, {\em Cambridge} 2007

\bibitem{BoOw10}
N. Bou-Rabee and H. Owhadi, Long-run accuracy of variational integrations in the stochastic context, {\em SIAM J. NUMER. ANAL.} {\textbf 48}, No. 1, 278-297 

\bibitem{Br14}
M. Bramanti, An Invitation to Hypoelliptic Operators and Hormander's Vector Field, {\em Springer} 2014



\bibitem{CaDe11}
M. Cadoni, R. De Leo, and Giuseppe, Composite 
model for DNA torsional dynamics, {\em Phys. Rev. E}, {\textbf 75}, 021919 (2007)

\bibitem{CaZh17}
G. Cai and W. Zhu, Elements of Stochastic Dynamics, {\em World Scientific} (2017)

\bibitem{ChTa22}
Y. Chao and M. Tao, Parametric resonance for enhancing the rate of metastable transition,
{\em SIAM J. APPL. MATH}, {\textbf 82}, 3, 1068-1090 



\bibitem{DaPe93}
T. Dauxois, M. Peyard, and A. R. Bishop,  Entropy-driven DNA denaturation, 
{\em Phys. Rev. E},  {\textbf 47} (1993) R44 

\bibitem{DeDe08}
R. De Leo and S. Demelio, Numerical analysis of solitons profiles in a composite model for DNA torsion dynamics, {\em Int. J. of Non-Linear Mechanics}, {\textbf 43} (2008) 1029-1039

\bibitem{DeZh07a}
M. Deng and W. Zhu, Energy diffusion controlled reaction rate in dissipative Hamiltonian systems, {\em Chinese Physics}, {\textbf 16} No. 6 (2007)

\bibitem{DeZh07b}
M. Deng and W. Zhu, On the stochastic dynamics of molecular conformation, {J. Zhejing Univ. Sci. A} (2007) 1401-1407

\bibitem{DuMe09}
P. Du Toit, I. Mezic, and  J. Marsden,  Coupled oscillator models with no scale separation, {\em Physica D}, \textbf{238}, 490 (2009).

\bibitem{DyGo01}
M. Dykman, B. Golding, L. Moccann, V. Smelyanskiy,  D. Luchinsky, R. Mannella, and P. McClintook, Activated escape of periodically driven systems, {\em Chaos}, {\textbf 11} (2001), 587-594

\bibitem{DyRa97}
M. Dykman, H. Rabitz, V. Smelyanskiy,  and H. Vugmeister, Resonant directed diffusion  in 
nonadiabatically driven systems, {\em Phys. Rev. Lett.}, {\textbf 79} (1997) 1178-1181

\bibitem{EiMe10} B. Eisenhower and I. Mezic, Targeted activation in deterministic and stochastic systems, {\em Phys. Rev. E}, \textbf{81}, 026603 (2010).

\bibitem{EnCa80}
Englander S. W., Calhoun D. B., Englander J. J., Kallenbach N. R., Liem R. K., Malin E. L., Mandal C., Rogero J. R., Individual breathing reactions measured in hemoglobin by hydrogen exchange methods, {\em Biophys J}. 1980 Oct, 32(1), 577-589.

\bibitem{Fr98}
M. I. Freidlin, Random and deterministic perturbations of nonlinear oscillators,
{\em Doc. Math. J.. DMV} (1998) 223-235

\bibitem{FrWeb98}
M. I. Freidlin and M. Weber, Random perturbations of nonlinear oscillators, {\em Ann. Probab.}, {\textbf 26}, No. 3 (1998) 925-967

\bibitem{FrWeb01}
M. I. Freidlin and M. Weber, On random perturbations of hamiltonian systems with many degrees of freedom, {\em Stochastic Processes and their Appllications },  {\textbf 94},  (2001) 199-239

\bibitem{FrWen12}
M. I. Freidlin and A. D. Wentzell, Random Perturbations of Dynamical Systems, {\em Springer}, 2012

\bibitem{GaKo05}
Gabern, F., W.~S. Koon, J.~E. Marsden, and S.~D. Ross [2005], Theory and
  computation of non-RRKM lifetime distributions and rates in chemical systems
  with three or more degrees of freedom, {\em Physica D} \textbf{211},
  391--406.

\bibitem{HoPu21}
C. M. Hough, D. N. Purschke, C. Huang, L. V. Titova, O. V. Kovalchuk, B. J. Warkentin, and F. A. Hegmann, Intense terahertz pulses inhibit Ras signaling and other cancer-associated signaling 
pathways in human skin tissue models, {\em J. Phys.,: Photonics} {\textbf 2021}, 3, 034004 

\bibitem{KoOw13} W. S. Koon, H. Owhadi, M. Tao, and T. Yanao, Control of a model of DNA division via parametric resonance, {\em Chaos}, {\textbf 23} 01317 (2013)

\bibitem{KrWi10}
T. Kristensen, W. Withayachumnankul, and P. Uhd Jeosen, and D. Abbott, Modeling terahertz
heating effects on water, {\em Optics Express} {\textbf 18}, No. 5 (2010)

\bibitem{LeMa15}
B. Leimkuhler and C. Matthews, Molecular Dynamics, with Deterministic and Stochastic Numerical Methods, {\em Springer} 2015 

\bibitem{MaMi22}
A. G. Markelz and D. M. Mittleman, Perspective on terahertz applications in bioscience and biotechnology, {\em ACS Photonics} 2022, 9, 1117-1126

\bibitem{Me06}
I. Mezic, On the dynamics of molecular conformation, {\em Proc. Natl. Acad. Sci,} 103
(20) (2006) 7542-7547.

\bibitem{Ot96}
H. C. Ottinger, Stochastic Processes in Polymeric Fluids. Tools and Examples for Developing Simulation Algorithms, {\em Springer} 1996

\bibitem{Pa14}
G. A. Pavliotis, Stochastic Processes and Applications, Diffusion Processes, the Fokker-Planck and Langevin Equations, {\em Springer} 2014

\bibitem{PeBi89}
M. Peyard and A. R. Bishop,  
Statistical mechanics of a nonlinear model for DNA denaturation, {\em Phys. Rev. Lett.}, {\textbf 62} (1989) 2755*

\bibitem{Ri72}
F. Riahi, On Lagrangians with high order derivatives, {\em Ann. J. Phys.}, {\textbf 40} (1972), 386-390

\bibitem{SiIl21}
D. M Sitnikov, I. V. Ilina, V. A. Revkova, A. Rodionov, S. A. Gurova, R. O. Shatalova, A. V. Kovalev, A. V. Ovchinnikov, O. V. Chefonov, M. A. Konoplyannikov, V. A. Kalsin, and V. P. Baklaushev, Effects of hight intensity non-ionizing terahertz radiation on human skin fibroblasts, 
{\em Biomed. Opt. Express}
{\textbf 2021}, 12, 7122-7138

\bibitem{SiRa13}
 I. Sizov, M. Rahman, B. Gelmont, M. Norton, and T. Globus, Sub-Thz spectroscopic characterization of  vibrational modes in artificially designed DNA monocrystal, {\em Chemical Physics} 425 (2013) 121-125

\bibitem{SmDy97}
V. Smelyanskiy, M. Dykman, H. Rabitz, and H. Vugmeister, Fluctuation, escape, and nucleation   in driven systems, {\em Phys. Rev. Lett.}, {\textbf 79} (1997) 3113-3116

\bibitem{SmMc99}
V. Smelyanskiy, P. Mcclintook,  R. Mannella, D. Luchinsky,  and M. Dykman, 
Thermally activated escape of driven systems: the activation energy {\em J. Phys. A Math. Gen.}, {\textbf 32} (1999)
 
\bibitem{SoTa19}
A. Souza and M. Tao, Metastable transition in inertial Langevin systems: what can be different from the overdamped case? {\em European J. Appl. Math.}, {\textbf 30} (2019) 830-852

\bibitem{StVa97}
D. W. Stroock and S. R. S. Varadhan, Multidimensional Diffusion Processes,
{\em Springer} 1997

\bibitem{Sw11}
E. S. Swanson, Modeling DNA response to terahertz radiation, {\em Phys. Rev. E} {\textbf 83} (2011)

\bibitem{TiAy13a}
L. V. Titova,  A. K. Ayesheshim, A. Golubov,  D. Fogen, R. Rodriguez-Juarez, R. Woycicki, F. A. Hegmann, and O. Kovalchuk, Intense THz pulses cause H2AX phosphorylation and activate DNA damage response in human skin tissue, {\em Biomed. Opt. Express} {\textbf 4} (2013) 

\bibitem{TiAy13b}
L. V. Titova,  A. K. Ayesheshim, A. Golubov,  R. Rodriguez-Juarez,  A. Kovalchuk, F. A. Hegmann, and O. Kovalchuk, Intense picosecond THz pulses alter gene expression in human skin tissue in vivo, {\em Proc. of SPIE}, {\textbf 8585} (2013)

\bibitem{TiAy13c}
L. V. Titova,  A. K. Ayesheshim, A. Golubov, R. Rodriguez-Juarez, R. Woycicki, F. A. Hegmann, and O. Kovalchuk, Intense THz pulses down-regulate genes associated with skin cancer and psoriasis: a new therapeutic avenue? {\em Sci. Rep.} 2013, 3, 2363

\bibitem{Ya04}
L. Yakushevich, {\em Nonlinear Physics of DNA}, Wiley-VCH, 2004.

\bibitem{Yo83}
S. Yomosa, Soliton excitations in deoxyribonucleic acid (DNA) double helices, {\em Phys. Rev. A}, \textbf{27}, Number 4 (1983)

\bibitem{Zh89}
C.-T. Zhang, Harmonic and subharmonic resonances of microwave absorption in DNA, {\em Phys. Rev. A} \textbf{40}, 2148-2153 (1989).

\end{thebibliography}
\end{document}